\documentclass[12pt,pdflatex]{elsarticle} 

\usepackage{natbib}

\usepackage[default,osfigures,scale=0.95]{opensans}
\usepackage[T1]{fontenc}
\usepackage{ae}
\usepackage[pdftex,hyperref,x11names]{xcolor}
\usepackage[pdftex]{hyperref}

\usepackage{blindtext}

\usepackage{geometry}
\usepackage{parallel}
\usepackage{parcolumns}
\usepackage{fancyhdr}

\usepackage{array}
\usepackage{amsmath}
\usepackage{amssymb}
\usepackage{amsfonts}
\usepackage{relsize}
\usepackage{mathtools}
\usepackage{bm}
\usepackage[%
decimalsymbol=.,
digitsep=fullstop
]{siunitx}

\usepackage[section]{placeins}

\usepackage{tabularx}
\usepackage{booktabs}
\usepackage{multicol}
\usepackage{multirow}
\usepackage{longtable}

\usepackage[font={small},labelfont=bf]{caption}

\usepackage[bottom]{footmisc}
   \newlength{\myfootnotesep}
\usepackage{enumerate}

\usepackage{paralist}

\usepackage{subfig}
\usepackage{graphicx}
\usepackage[space]{grffile} 
\usepackage{placeins} 
\usepackage{tikz}
\usepackage{rotating}

\usepackage[doublespacing]{setspace}

\newcommand{\pkg}[1]{{\fontseries{b}\selectfont #1}}

\newcommand\simiid{\stackrel{\mathclap{\normalfont\mbox{\tiny{iid}}}}{\sim}}

\makeatletter
\def\ps@pprintTitle{%
 \let\@oddhead\@empty
 \let\@evenhead\@empty
 \def\@oddfoot{}%
 \let\@evenfoot\@oddfoot}

\def\input@path{{/Users/janus829/Dropbox/Research/netModels/summResults/}, {/Users/s7m/Dropbox/Research/netModels/summResults/}, {/Users/mdw/Dropbox/netModels/summResults/}}
\graphicspath{{/Users/janus829/Dropbox/Research/netModels/summResults/}, {/Users/s7m/Dropbox/Research/netModels/summResults/},{/Users/mdw/Dropbox/netModels/summResults/}}
\makeatother

\begin{document}

\thispagestyle{empty}
\begin{frontmatter}

\title{Inferential Approaches for Network Analysis: \\ AMEN for Latent Factor Models\tnoteref{t1} \\ \textit{Forthcoming Political Analysis}}

\tnotetext[t1]{Shahryar Minhas and Michael D. Ward acknowledge support from National Science Foundation (NSF) Award 1259266 and Peter D. Hoff acknowledges support from NSF Award 1505136. Replication files for this project can be accessed at \url{https://github.com/s7minhas/netmodels}.}

\author[msu]{Shahryar Minhas\corref{cor1}}
\ead{minhassh@msu.edu}
\cortext[cor1]{Corresponding author}
\author[duke2]{Peter D. Hoff}
\author[duke]{Michael D. Ward}

\address[msu]{Department of Political Science, Michigan State University, East Lansing, MI 48824, USA}
\address[duke]{Department of Political Science, Duke University, Durham, NC 27701, USA}
\address[duke2]{Department of Statistical Science, Duke University, Durham, NC 27701, USA}

\begin{abstract}
	\singlespacing{
		We introduce a Bayesian approach to conduct inferential analyses on dyadic data while accounting for interdependencies between observations through a set of additive and multiplicative effects (AME). The AME model is built on a generalized linear modeling framework and is thus flexible enough to be applied to a variety of contexts. We contrast the AME model to two prominent approaches in the literature: the latent space model (LSM) and the exponential random graph model (ERGM).  Relative to these approaches, we show that the AME approach is a) to be easy to implement; b) interpretable in a general linear model framework; c) computationally straightforward; d) not prone to degeneracy; e) captures 1st, 2nd, and 3rd order network dependencies; and f) notably outperforms ERGMs and LSMs on a variety of metrics and in an out-of-sample context. In summary, AME offers a straightforward way to undertake nuanced, principled inferential network analysis for a wide range of social science questions.
		}
\end{abstract}
\end{frontmatter}


\newpage\setcounter{page}{1} 

Data structures that define relations between pairs of actors are ubiquitous in political science -- examples include the study of events such as legislation cosponsorship, trade, interstate conflict, and the formation of international agreements. The dominant paradigm for dealing with such data, however, is not a network approach but rather a dyadic design, in which an interaction between a pair of actors is considered independent of interactions between any other pair in the system. To highlight the ubiquity of this approach the following represent just a sampling of the articles published from the 1980s to the present in the American Journal of Political Science (AJPS) and American Political Science Review (APSR) that assume dyadic independence: \citet{dixon:1983,mansfield:etal:2000,lemke:reed:2001a,mitchell:2002,dafoe:2011a,fuhrmann:sechser:2014,carnegie:2014}.

The implication of this assumption is that when, for example, Vietnam and the United States decide to form a trade agreement, they make this decision independently of what they have done with other countries and what other countries in the international system have done among themselves.\footnote{There has been significant work done on treaty formation that would challenge this claim, e.g., see \citet{manger:etal:2012,kinne:2013}.} An even stronger assumption is that Japan declaring war against the United States is independent of the decision of the United States to go to war against Japan.\footnote{\citet{maoz:etal:2006,minhas:etal:2016} would each note the importance of taking into account network dynamics in the study of interstate conflict.} A common refrain from those that favor the dyadic approach is that many events are only bilateral (\citealt{diehl:wright:2016}), thus alleviating the need for an approach that incorporates interdependencies between observations. However, even bilateral events and processes take place within a broader system, and occurrences in one part of the system may be dependent upon events in another. At a minimum, we do not know whether independence of events and processes characterizes what we observe. 

In this article, we introduce the additive and multiplicative effects (AME) model and compare it to two popular alternatives: the latent space model (LSM) and exponential random graph model (ERGM). The AME approach to network modeling is a flexible framework that can be used to estimate many different types of cross-sectional and longitudinal networks with binary, ordinal, or continuous edges within a generalized linear model framework. Our approach addresses ways in which observations can be interdependent while still allowing scholars to focus on examining theories that may only be relevant in the monadic or dyadic level. Further, at the network level it accounts for nodal and dyadic dependence patterns, and provides a descriptive visualization of higher-order dependencies such as homophily and stochastic equivalence. 

The article is organized as follows, we begin by briefly discussing the difficulties in studying dyadic data through approaches that assume observational independence. Then we introduce the AME framework in two steps. We first discuss nodal and dyadic dependencies that may lead to non-iid observations and show how the additive effects portion of AME can be used to model these dependencies. Similarly, in the second step, we discuss how the multiplicative effects portion of the AME framework can be used to effectively model third order effects while still enabling researchers to study exogenous covariates of interest. We then briefly contrast these latent variables models with ERGM and conclude with an application on a cross-sectional network measuring collaborations during the policy design of the Swiss CO$_{2}$ act. We show that AME provides a superior goodness of fit to the data in terms of ability to predict linkages and capture network dependencies. 
\\

\section*{\textbf{Addressing Dependencies in Dyadic Data}}

Relational, or dyadic, data provide measurements of how pairs of actors relate to one another. The easiest way to organize such data is the directed dyadic design in which the unit of analysis is some set of $n$ actors that have been paired together to form a dataset of $z$ directed dyads. A tabular design such as this for a set of $n$ actors, $\{i, j, k, l \}$ results in $n \times (n-1)$ observations, as shown in Table~\ref{tab:canDesign}. 

\begin{table}[ht]
	\captionsetup{justification=raggedright }
	\centering
	\begin{minipage}{.45\textwidth}
		\centering
		\begingroup
		\setlength{\tabcolsep}{10pt}
		\begin{tabular}{ccc}
			Sender & Receiver & Event \\
			\hline\hline
			$i$ & $j$ & $y_{ij}$ \\
			\multirow{2}{*}{\vdots} & $k$ & $y_{ik}$ \\
			~ & $l$ & $y_{il}$ \\
			$j$ & $i$ & $y_{ji}$ \\
			\multirow{2}{*}{\vdots} & $k$ & $y_{jk}$ \\
			~ & $l$ & $y_{jl}$ \\
			$k$ & $i$ & $y_{ki}$ \\
			\multirow{2}{*}{\vdots} & $j$ & $y_{kj}$ \\
			~ & $l$ & $y_{kl}$ \\
			$l$ & $i$ & $y_{li}$ \\
			\multirow{2}{*}{\vdots} & $j$ & $y_{lj}$ \\
			~ & $k$ & $y_{lk}$ \\
			\hline\hline
		\end{tabular}
		\endgroup
		\caption{Structure of datasets used in canonical design.} 
		\label{tab:canDesign}
	\end{minipage}
	$\mathbf{\longrightarrow}$
	\begin{minipage}{.45\textwidth}
		\centering
		\begingroup
		\setlength{\tabcolsep}{10pt}
		\renewcommand{\arraystretch}{1.5}
		\begin{tabular}{c||cccc}
		~ & $i$ & $j$ & $k$ & $l$ \\ \hline\hline
		$i$ & \footnotesize{NA} & $y_{ij}$ & $y_{ik}$ & $y_{il}$ \\
		$j$ & $y_{ji}$ & \footnotesize{NA}  & $y_{jk}$ & $y_{jl}$ \\
		$k$ & $y_{ki}$ & $y_{kj}$ & \footnotesize{NA}  & $y_{kl}$ \\
		$l$ & $y_{li}$ & $y_{lj}$ & $y_{lk}$ & \footnotesize{NA}  \\
		\end{tabular}
		\endgroup
		\caption{Adjacency matrix representation of data in Table~\ref{tab:canDesign}. Senders are represented by the rows and receivers by the columns. }
		\label{tab:netDesign}
	\end{minipage}
\end{table}

When modeling these types of data, scholars typically employ a generalized linear model (GLM) estimated via maximum-likelihood. The stochastic component of this model reflects our assumptions about the probability distribution from which the data are generated: $y_{ij} \sim P(Y | \theta_{ij})$, with a probability density or mass function such as the normal, binomial, or Poisson, and we assume that each dyad in the sample is independently drawn from that particular distribution, given $\theta_{ij}$. The systematic component characterizes the model for the parameters of that distribution and describes how $\theta_{ij}$ varies as a function of a set of nodal and dyadic covariates, $\mathbf{X}_{ij}$: $\theta_{ij} = \bm\beta^{T} \mathbf{X}_{ij}$. The key assumption we make when applying this modeling technique is that given $\mathbf{X}_{ij}$ and the parameters of our distribution, each of the dyadic observations is conditionally independent. Specifically, we construct the joint density function over all dyads using the observations from Table 1 as an example.

\vspace{-8mm}
\begin{align}
\begin{aligned}
	P(y_{ij}, y_{ik}, \ldots, y_{lk} | \theta_{ij}, \theta_{ik}, \ldots, \theta_{lk}) &= P(y_{ij} | \theta_{ij}) \times P(y_{ik} | \theta_{ik}) \times \ldots \times P(y_{lk} | \theta_{lk}) \\
	P(\mathbf{Y} \; | \; \bm{\theta}) &= \prod_{\alpha=1}^{n \times (n-1)} P(y_{\alpha} | \theta_{\alpha})  \\
\end{aligned}
\end{align}

\noindent We next convert the joint probability into a likelihood: $\displaystyle \mathcal{L} (\bm{\theta} | \mathbf{Y}) = \prod_{\alpha=1}^{n \times (n-1)} P(y_{\alpha} | \theta_{\alpha})$.

The likelihood as defined above is only valid if we are able to make the assumption that, for example, $y_{ij}$ is independent of $y_{ji}$ and $y_{ik}$ given the set of covariates we specified.\footnote{The difficulties of applying the GLM framework to data that have structural interdependencies between observations is a problem that has long been recognized. \citet{beck:katz:1995}, for example, detail the issues with pooling observations in time-series cross-section datasets.} Assuming that the dyad $y_{ij}$ is conditionally independent of the dyad $y_{ji}$ asserts that there is no level of reciprocity in a dataset, an assumption that in many cases would seem quite untenable. A harder problem to handle is the assumption that $y_{ij}$ is conditionally independent of $y_{ik}$, the difficulty here follows from the possibility that $i$'s relationship with $k$ is dependent on how $i$ relates to $j$ and how $j$ relates to $k$, or more simply put the ``enemy of my enemy [may be] my friend''. Accordingly, inferences drawn from misspecified models that ignore potential interdependencies between dyadic observations are likely to have a number of issues including biased estimates of the effect of independent variables, uncalibrated confidence intervals, and poor predictive performance.


\section*{\textbf{Additive Part of AME}}

The dependencies that tend to develop in relational data can be more easily understood when we move away from stacking dyads on top of one another and turn instead to a matrix design as illustrated in Table~\ref{tab:netDesign}. Operationally, this type of data structure is represented as a $n \times n$ matrix, $\mathbf{Y}$, where the diagonals are typically undefined. The $ij^{th}$ entry defines the relationship sent from $i$ to $j$ and can be continuous or discrete. Relations between actors in a network setting at times does not involve senders and receivers. Networks such as these are referred to as undirected and all the relations between actors are symmetric, meaning $y_{ij}=y_{ji}$.

The most common type of dependency that arises in networks are first-order, or nodal dependencies, and these point to the fact that we typically find significant heterogeneity in activity levels across nodes. The implication of this across-node heterogeneity is within-node homogeneity of ties, meaning that values across a row, say $\{y_{ij},y_{ik},y_{il}\}$, will be more similar to each other than other values in the adjacency matrix because each of these values has a common sender $i$. This type of dependency manifests in cases where sender $i$ tends to be more active or less active in the network than other senders. Similarly, while some actors may be more active in sending ties to others in the network, we might also observe that others are more popular targets, this would manifest in observations down a column, $\{y_{ji},y_{ki},y_{li}\}$, being more similar. Last, we might also find that actors who are more likely to send ties in a network are also more likely to receive them, meaning that the row and column means of an adjacency matrix may be correlated. Another ubiquitous type of structural interdependency is reciprocity. This is a second-order, or dyadic, dependency relevant only to directed datasets, and asserts that values of $y_{ij}$ and $y_{ji}$ may be statistically dependent. The prevalence of these types of potential interactions within directed dyadic data also complicates the basic assumption of observational independence.
 
We model first- and second-order dependencies in AME using a set of additive effects that are motivated by the social relations model (SRM) developed by \citep{warner:etal:1979,li:loken:2002}. Specifically, we decompose the variance of observations in an adjacency matrix in terms of heterogeneity across row means (out-degree), heterogeneity along column means (in-degree), correlation between row and column means, and correlations within dyads:

\begin{align}
\begin{aligned}
	y_{ij} &= \mu + e_{ij} \\
	e_{ij} &= a_{i} + b_{j} + \epsilon_{ij} \\
	\{ (a_{1}, b_{1}), \ldots, (a_{n}, b_{n}) \} &\simiid N(0,\Sigma_{ab}) \\ 
	\{ (\epsilon_{ij}, \epsilon_{ji}) : \; i \neq j\} &\simiid N(0,\Sigma_{\epsilon}), \text{ where } \\
	\Sigma_{ab} = \begin{pmatrix} \sigma_{a}^{2} & \sigma_{ab} \\ \sigma_{ab} & \sigma_{b}^2   \end{pmatrix} \;\;\;\;\; &\Sigma_{\epsilon} = \sigma_{\epsilon}^{2} \begin{pmatrix} 1 & \rho \\ \rho & 1  \end{pmatrix} .
\label{eqn:srmCov}
\end{aligned}
\end{align}

$\mu$ here provides a baseline measure of the density mean of a network, and $e_{ij}$ represents residual variation. The residual variation decomposes into parts: a row/sender effect ($a_{i}$), a column/receiver effect ($b_{j}$), and a within-dyad effect ($\epsilon_{ij}$). The row and column effects are modeled jointly to account for correlation in how active an actor is in sending and receiving ties. Heterogeneity in the row and column means is captured by $\sigma_{a}^{2}$ and $\sigma_{b}^{2}$, respectively, and $\sigma_{ab}$ describes the linear relationship between these two effects (i.e., whether actors who send  a lot of ties also receive  a lot of ties). Beyond these first-order dependencies, second-order dependencies are described by $\sigma_{\epsilon}^{2}$ and a within dyad correlation, or reciprocity, parameter $\rho$. 

We incorporate the covariance structure described in Equation~\ref{eqn:srmCov} into the systematic component of a GLM framework: $\bm\beta^{\top} \mathbf{X}_{ij} + a_{i} + b_{j} + \epsilon_{ij}$, where $ \bm\beta^{\top} \mathbf{X}_{ij}$ accommodates the inclusion of dyadic, sender, and receiver covariates. This approach incorporates row, column, and within-dyad dependence in way that is widely used and understood by applied researchers: a regression framework and additive random effects to accommodate variances and covariances often seen in relational data. Furthermore, this handles a diversity of outcome distributions. 

\section*{Multiplicative Part of AME}

Missing from the additive effects portion of the model is an accounting of third-order dependence patterns that can arise in relational data. A third-order dependency is defined as the dependency between triads, not dyads. The ubiquity of third-order effects in relational datasets can arise from the presence of some set of shared attributes between nodes that affects their probability of interacting with one another.\footnote{Another reason why we may see the emergence of third-order effects is the ``sociology'' explanation: that individuals want to close triads because this is putatively a more stable or preferable social situation (\citealt{wasserman:faust:1994}).}

\begin{figure}[ht]
	\centering
	\caption{Graph on the left is a representation of an undirected network that exhibits a high degree of homophily (linkages forming because of shared attributes), while on the right we show an undirected network that exhibits stochastic equivalence.}	
	\begin{tabular}{lcr}
	\includegraphics[width=.33\textwidth]{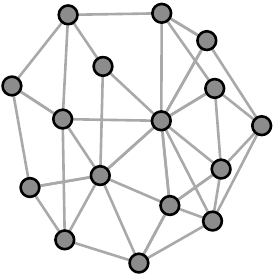} & \hspace{2cm} &
	\includegraphics[width=.33\textwidth]{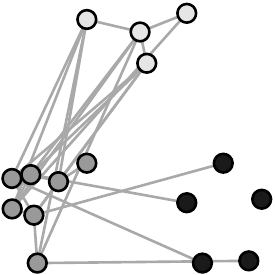}	
	\end{tabular}
	\label{fig:homphStochEquivNet}
\end{figure}

For example, finding common in the political economy literature is that democracies are more likely to form trade agreements with one another, and the shared attribute here is a country's political system. A binary network where actors tend to form ties with others based on some set of shared characteristics often leads to a network graph with a high number of ``transitive triads'' in which  sets of actors $\{i,j,k\}$ are each linked to one another. The left-most plot in Figure~\ref{fig:homphStochEquivNet} provides a representation of a network that exhibits this type of pattern. The relevant implication of this when it comes to conducting statistical inference is that--unless we are able to specify the list of exogenous variable that may explain this prevalence of triads--the probability of $j$ and $k$ forming a tie is not independent of the ties that already exist between those actors and $i$.

Another third-order dependence pattern that cannot be accounted for in the additive effects framework is stochastic equivalence. A pair of actors $ij$ are stochastically equivalent if the probability of $i$ relating to, and being related to, by every other actor is the same as the probability for $j$. This refers to the idea that there will be groups of nodes in a network with similar relational patterns. The occurrence of a dependence pattern such as this is not uncommon in the social science applications. Recent work estimates a stochastic equivalence structure to explain the formation of preferential trade agreements (PTAs) between countries \cite{manger:etal:2012}. Specifically, they suggest that PTA formation is related to differences in per capita income levels between countries. Countries falling into high, middle, and low income per capita levels will have patterns of PTA formation that are determined by the groups into which they fall. Such a structure is represented in the right-most panel of Figure~\ref{fig:homphStochEquivNet}, here the lightly shaded group of nodes at the top can represent high-income countries, nodes on the bottom-left middle-income, and the darkest shade of nodes low-income countries. The behavior of actors in a network can at times be governed by group level dynamics, and failing to account for such dynamics leaves potentially important parts of the data generating process ignored.

We account for third order dependence patterns using a latent variable framework, and our goal in doing so is twofold: 1) be able to adequately represent third order dependence patterns, 2) improve our ability to conduct inference on exogenous covariates. Latent variable models assume that relationships between nodes are mediated by a small number ($K$) of node-specific unobserved latent variables. We contrast the approach that we utilize within AME, the latent factor model (LFM), to the latent space model, which is among the most widely used in the networks literature.\footnote{An alternative approach with a similar latent variable formulation is known as the stochastic block model \citep{nowicki:snijders:2001}, however, this approach is typically only used to model community structure in networks and not used to conduct inference on exogenous covariates.} For the sake of exposition, we consider the case where relations are symmetric to describe the differences between these approaches. These approaches can be incorporated into the framework that we have been constructing through the inclusion of an additional term, $\alpha(\mu_{i}, \mu_{j})$, that captures latent third order characteristics of a network. General definitions for how $\alpha(u_{i}, u_{j})$ are defined for these latent variable models are shown in Equations~\ref{eqn:latAlpha}:

\begin{align}
\begin{aligned}
\text{Latent space model} \\
	&\alpha(\textbf{u}_{i}, \textbf{u}_{j}) = -|\textbf{u}_{i} - \textbf{u}_{j}| \\
	&\textbf{u}_{i} \in \mathbb{R}^{K}, \; i \in \{1, \ldots, n \} \\
\text{Latent factor model} \\
	&\alpha(\textbf{u}_{i}, \textbf{u}_{j}) = \textbf{u}_{i}^{\top} \Lambda \textbf{u}_{j} \\
	&\textbf{u}_{i} \in \mathbb{R}^{K}, \; i \in \{1, \ldots, n \} \\
	&\Lambda \text{ a } K \times K \text{ diagonal matrix}
\label{eqn:latAlpha}
\end{aligned}
\end{align}

In the LSM approach, each node $i$ has some unknown latent position in $K$ dimensional space, $\textbf{u}_{i} \in \mathbb{R}^{K}$, and the probability of a tie between a pair $ij$ is a function of the negative Euclidean distance between them: $-|\textbf{u}_{i} - \textbf{u}_{j}|$. Because latent distances for a triple of actors obey the triangle inequality, this formulation models the tendencies toward homophily commonly found in social networks. This approach is implemented in the \pkg{latentnet} which is part of the \pkg{statnet} $\sf{R}$ package \cite{krivitsky:handcock:2015}. However, this approach also comes with an important shortcoming: it confounds stochastic equivalence and homophily. Consider two nodes $i$ and $j$ that are proximate to one another in $K$ dimensional Euclidean space, this suggests not only that $|\textbf{u}_{i} - \textbf{u}_{j}|$ is small but also that $|\textbf{u}_{i} - \textbf{u}_{l}| \approx |\textbf{u}_{j} - \textbf{u}_{l}|$, the result being that nodes $i$ and $j$ will by construction assumed to possess the same relational patterns with other actors such as $l$ (i.e., that they are stochastically equivalent). Thus LSMs confound strong ties with stochastic equivalence. This approach cannot adequately model data with many ties between nodes that have different network roles. This is problematic as real-world networks exhibit varying degrees of stochastic equivalence and homophily. In these situations, using the LSM would end up representing only a part of the network structure. 

In the latent factor model, each actor has an unobserved vector of characteristics, $\textbf{u}_{i} = \{u_{i,1}, \ldots, u_{i,K} \}$, which describe their behavior as an actor in the network. The probability of a tie from $i$ to $j$ depends on the extent to which $\textbf{u}_{i}$ and $\textbf{u}_{j}$ are ``similar'' (i.e., point in the same direction) and on whether the entries of $\Lambda$ are greater or less than zero. More specifically, the similarity in the latent factors, $\textbf{u}_{i} \approx \textbf{u}_{j}$, corresponds to how stochastically equivalent a pair of actors are and the eigenvalue determines whether the network exhibits positive or negative homophily. For example, say that we estimate a rank-one latent factor model (i.e., $K=1$), in this case $\textbf{u}_{i}$ is represented by a scalar $u_{i,1}$, similarly, $\textbf{u}_{j}=u_{j,1}$, and $\Lambda$ will have just one diagonal element $\lambda$. The average effect this will have on $y_{ij}$ is simply $\lambda \times u_{i} \times u_{j}$, where a positive value of $\lambda>0$ indicates homophily and $\lambda<0$ heterophily. This approach can represent both varying degrees of homophily and stochastic equivalence.\footnote{In the directed version of this approach, we use the singular value decomposition, here actors in the network have a vector of latent characteristics to describe their behavior as a sender, denoted by $\textbf{u}$, and as a receiver, $\textbf{v}$: $\textbf{u}_{i}, \textbf{v}_{j} \in \mathbb{R}^{K}$. This can alter the probability of an interaction between $ij$ additively: $\textbf{u}_{i}^{\top} \textbf{D} \textbf{v}_{j}$, where $\textbf{D}$ is a $K \times K$ diagonal matrix.}

In addition to summarizing dependence patterns in networks, scholars are often concerned with accounting for interdependencies so that they can better estimate the effects of exogenous covariates. Both the latent space and factor models attempt to do this as they are ``conditional independence models'' --  in that they assume that ties are conditionally independent given all of the observed predictors and unknown node-specific parameters: $p( Y | X , U ) = \prod_{i<j} p( y_{i,j}  | x_{i,j} , u_i , u_j)$. Typical parametric models of this form relate $y_{i,j}$ to $(x_{i,j},u_i,u_j)$ via a link function:

\begin{align*}
	p(y_{i,j} | x_{i,j}, u_i , u_j ) & = f( y_{i,j} : \eta_{i,j} ) \\
	\eta_{i,j} &= \beta^\top x_{i,j} + \alpha(\textbf{u}_{i}, \textbf{u}_{j}).
\end{align*}

However, the structure of $\alpha(\textbf{u}_{i}, \textbf{u}_{j})$ can result in very different interpretations for any estimates of the regression coefficients $\beta$. For example, suppose the latent effects $\{ u_1,\ldots, u_n\}$ are near zero on average (if not, their mean can be absorbed into an intercept parameter and row and column additive effects). Under the LFM, the average value of $\alpha(\textbf{u}_{i}, \textbf{u}_{j}) = \textbf{u}_{i}^{\top} \Lambda \textbf{u}_{j}$ will be near zero and so we have

\begin{align*}
	\eta_{i,j} & =  \beta^\top x_{i,j} + \textbf{u}_{i}^{\top} \Lambda \textbf{u}_{j} \\
	\bar \eta & \approx  \beta^\top \bar x.
\end{align*}

The implication of this is that the values of $\beta$ can be interpreted as yielding the ``average'' value of $\eta_{i,j}$. On the other hand, under the LSM

\begin{align*}
	\eta_{i,j} & =  \beta^\top x_{i,j} - |\textbf{u}_{i} - \textbf{u}_{j}|  \\
	\bar \eta & \approx  \beta^\top \bar x - \overline{ |\textbf{u}_{i} - \textbf{u}_{j}| } <  \beta^\top \bar x .
\end{align*}

In this case, $\beta^\top \bar x$ does not represent an ``average'' value of the predictor $\eta_{i,j}$, it represents a maximal value as if all actors were zero distance from each other in the latent social space. For example, consider the simplest case of a normally distributed network  outcome with an identity link:

\begin{align*}
	y_{i,j} & = \beta^\top x_{i,j} + \alpha(\textbf{u}_{i}, \textbf{u}_{j}) + \epsilon_{i,j} \\
	\bar y & \approx \beta^\top \bar x + \overline{ \alpha(\textbf{u}_{i}, \textbf{u}_{j}) }   .
\end{align*}

Under the LSM, $\bar y \approx \beta^\top \bar x - \overline{ |\textbf{u}_{i} - \textbf{u}_{j}|  } < \beta^\top \bar x$, and so we no longer can interpret $\beta$ as representing the linear relationship between $y$ and $x$. Instead, it may be thought of as describing some sort of average hypothetical ``maximal'' relationship between $y_{i,j}$ and $x_{i,j}$.

Thus the LFM provides two important benefits. First, we are able to capture a wider assortment of dependence patterns that arise in relational data, and, second, parameter interpretation is more straightforward. The AME approach considers the regression model shown in Equation~\ref{eqn:ame}:

\begin{align}
\begin{aligned}
	y_{ij} &= g(\theta_{ij}) \\
	&\theta_{ij} = \bm\beta^{\top} \mathbf{X}_{ij} + e_{ij} \\
	&e_{ij} = a_{i} + b_{j}  + \epsilon_{ij} + \alpha(\textbf{u}_{i}, \textbf{v}_{j}) \text{  , where } \\
	&\qquad \alpha(\textbf{u}_{i}, \textbf{v}_{j}) = \textbf{u}_{i}^{\top} \textbf{D} \textbf{v}_{j} = \sum_{k \in K} d_{k} u_{ik} v_{jk}. \\
\label{eqn:ame}
\end{aligned}
\end{align}

Using this framework, we are able to model the dyadic observations as conditionally independent given $\bm\theta$, where $\bm\theta$ depends on the the unobserved random effects, $\mathbf{e}$. $\mathbf{e}$ is then modeled to account for the potential first, second, and third-order dependencies that we have discussed. As described in Equation~\ref{eqn:srmCov}, $a_{i} + b_{j}  + \epsilon_{ij}$, are the additive random effects in this framework and account for sender, receiver, and within-dyad dependence. The multiplicative effects, $\textbf{u}_{i}^{\top} \textbf{D} \textbf{v}_{j}$, are used to capture higher-order dependence patterns that are left over in $\bm\theta$ after accounting for any known covariate information.\footnote{The MCMC algorithm describing the estimation procedure is available in the Appendix.} 

\subsubsection*{\textbf{ERGMs}}

An alternative approach to accounting for third-order dependence patterns are ERGMs. Whereas AME seeks to estimate interdependencies in a network through a set of latent variables, ERGM approaches are useful when researchers are interested in the role that a specific network statistic(s) has in giving rise to an observed network. These network statistics could include the number of transitive triads in a network, balanced triads, reciprocal pairs and so on.\footnote{\citet{morris:etal:2008} and \citet{snijders:etal:2006} provide a detailed list of network statistics that can be included in an ERGM model specification.} In the ERGM framework, a set of statistics, $S(\mathbf{Y})$, define a model. Given the chosen set of statistics, the probability of observing a particular network dataset $\mathbf{Y}$ can be expressed as:

\begin{align}
\Pr(Y = y) = \frac{ \exp( \bm\beta^{T} S(y)  )  }{ \sum_{z \in \mathcal{Y}} \exp( \bm\beta^{T} S(z)  )  } \text{ ,  } y \in \mathcal{Y}
\label{eqn:ergm}
\end{align}

$\bm\beta$ represents a vector of model coefficients for the specified network statistics, $\mathcal{Y}$ denotes the set of all obtainable networks, and the denominator is used as a normalizing factor \citep{hunter:etal:2008}. This approach provides a way to state that the probability of observing a given network depends on the patterns that it exhibits, which are operationalized in the list of network statistics specified by the researcher. Within this approach one can test the role that a variety of network statistics play in giving rise to a particular network. Additionally, researchers can easily accommodate nodal and dyadic covariates. Further because of the Hammersley-Clifford theorem any probability distribution over networks can be represented by the form shown in Equation~\ref{eqn:ergm}. 

A notable issue when estimating ERGMs, however, is that the estimated model can become degenerate. Degeneracy here means that the model places a large amount of probability on a small subset of networks that fall in the set of obtainable networks, $\mathcal{Y}$, but share little resemblance with the observed network, $\mathbf{Y}$ \citep{schweinberger:2011}.\footnote{For example, most of the probability may be placed on empty graphs, no edges between nodes, or nearly complete graphs, almost every node is connected by an edge.} Some have argued that model degeneracy is simply a result of model misspecification \citep{handcock:2003a,goodreau:etal:2008,handcock:etal:2008}. However, this points to an important caveat in interpreting the implications of the Hammersley-Clifford theorem. Though this theorem ensures that any network can be represented through an ERGM, it says nothing about the complexity of the sufficient statistics ($S(y)$) required to do so. Failure to properly account for higher-order dependence structures through an appropriate specification can at best lead to model degeneracy, which provides an obvious indication that the specification needs to be altered, and at worst deliver a result that converges but does not appropriately capture the interdependencies in the network. The consequence of the latter case is a set of inferences that will continue to be biased as a result of unmeasured heterogeneity, thus defeating the major motivation for pursuing an inferential network model in the first place.

In the following section we undertake a comparison of the latent distance model, ERGM, and the AME model using an application presented in \citet{cranmer:etal:2016}.\footnote{The reason we use the same dataset is because of the model specification issue that arises when using ERGMs. As \citet[p. 8]{cranmer:etal:2016} note, when using ERGMs scholars must model third-order effects and ``must also specify them in a complete and correct manner'' or the model will be misspecified. Thus to avoid providing an incorrect specification when comparing ERGM with the AME we use the specification that they constructed.} In doing so, we are able to compare and contrast these various approaches.

\section*{\textbf{Empirical Comparison}}

We utilize a cross-sectional network measuring whether an actor indicated that they collaborated with each other during the policy design of the Swiss CO$_{2}$ act (\citealt{ingold:2008}). This is a directed relational matrix as an actor $i$ can indicate that they collaborated with $j$ but $j$ may not have stated that they collaborated with $i$. The Swiss government proposed this act in 1995 with the goal of undertaking a 10\% reduction in CO$_{2}$ emissions by 2012. The act was accepted in the Swiss Parliament in 2000 and implemented in 2008. \citet{ingold:2008}, and subsequent work by \citet{ingold:fischer:2014}, sought to determine what drives collaboration among actors trying to affect climate change policy. The set of actors included in this network are those that were identified by experts as holding an important position in Swiss climate policy. In total, Ingold identifies 34 relevant actors: five state actors, eleven industry and business representatives, seven environmental NGOs and civil society organizations, five political parties, and six scientific institutions and consultants. We follow Ingold \& Fischer and \citet{cranmer:etal:2016} in developing a model specification to understand and predict link formation in this network.\footnote{We do not review the specification in detail here, instead we just provide a summary of the variables to be included and the theoretical expectations of their effects in the Appendix.}

The LSM we fit on this network includes a two-dimensional Euclidean distance metric. The ERGM specification for this network includes the same exogenous variables as LSM, but also includes a number of endogenous characteristics of the network. The AME model we fit includes the same exogenous covariates and accounts for nodal and dyadic heterogeneity using the SRM.\footnote{Convergence diagnostics for AME are provided in the Appendix.} Third-order effects are represented by the latent factor model with $K=2$. Last, we also include a logistic model as that is still the standard in most of the field. Parameter estimates for these three approaches are shown in Table~\ref{tab:regTable}.

The first point to note is that, in general, the parameter estimates returned by the AME while similar to those of ERGM are quite different from the LSM. For example, while the LSM returns a result for the \texttt{Opposition/alliance} variable that diverges from ERGM, the AME returns a result that is similar to Ingold \& Fischer. Similar discrepancies appear for other parameters such as \texttt{Influence attribution} and \texttt{Alter's influence degree}. Each of these discrepancies are eliminated when using AME. As described previously, this is because the LSM approach complicates the interpretation of the effects of exogenous variables due to the construction of the latent variable term.\footnote{In the Appendix, we show that these differences persist even when incorporating sender and receiver random effects into the LSM.}

\begin{table}[ht]
\centering
\caption{Logit and ERGM results are shown with standard errors in parentheses. LSM and AME are shown with 95\% posterior credible intervals provided in brackets.}
\begin{tabular}{lcccc}
   & Logit & LSM & ERGM & AME \\ 
  \hline\hline
  Intercept/Edges & -4.44 & 0.95 & -12.17 & -3.40 \\ 
   & (0.34) & [0.09; 1.85] & (1.40) & [-4.40; -2.51] \\ 
  \textbf{Conflicting policy preferences} &  &  &  &  \\ 
  $\;\;\;\;$ Business vs. NGO & -0.86 & -1.37 & -1.11 & -1.38 \\ 
   & (0.46) & [-2.39; -0.40] & (0.51) & [-2.47; -0.49] \\ 
  $\;\;\;\;$ Opposition/alliance & 1.21 & 0.00 & 1.22 & 1.08 \\ 
   & (0.20) & [-0.40; 0.40] & (0.20) & [0.72; 1.49] \\ 
  $\;\;\;\;$ Preference dissimilarity & -0.07 & -1.77 & -0.44 & -0.79 \\ 
   & (0.37) & [-2.64; -0.91] & (0.39) & [-1.55; -0.07] \\ 
  \textbf{Transaction costs} &  &  &  &  \\ 
  $\;\;\;\;$ Joint forum participation & 0.88 & 1.51 & 0.90 & 0.92 \\ 
   & (0.27) & [0.85; 2.17] & (0.28) & [0.40; 1.46] \\ 
  \textbf{Influence} &  &  &  &  \\ 
  $\;\;\;\;$ Influence attribution & 1.20 & 0.08 & 1.00 & 1.10 \\ 
   & (0.22) & [-0.40; 0.54] & (0.21) & [0.70; 1.55] \\ 
  $\;\;\;\;$ Alter's influence indegree & 0.10 & 0.01 & 0.21 & 0.11 \\ 
   & (0.02) & [-0.03; 0.04] & (0.04) & [0.07; 0.15] \\ 
  $\;\;\;\;$ Influence absolute diff. & -0.03 & 0.04 & -0.05 & -0.07 \\ 
   & (0.02) & [-0.01; 0.09] & (0.01) & [-0.11; -0.03] \\ 
  $\;\;\;\;$ Alter = Government actor & 0.63 & -0.46 & 1.04 & 0.56 \\ 
   & (0.25) & [-1.08; 0.14] & (0.34) & [-0.06; 1.16] \\ 
  \textbf{Functional requirements} &  &  &  &  \\ 
  $\;\;\;\;$ Ego = Environmental NGO & 0.88 & -0.60 & 0.79 & 0.68 \\ 
   & (0.26) & [-1.30; 0.08] & (0.17) & [-0.36; 1.73] \\ 
  $\;\;\;\;$ Same actor type & 0.74 & 1.17 & 0.99 & 1.03 \\ 
   & (0.22) & [0.62; 1.72] & (0.23) & [0.62; 1.48] \\ 
  \textbf{Endogenous dependencies} &  &  &  &  \\ 
  $\;\;\;\;$ Mutuality & 1.22 &  & 0.81 &  \\ 
   & (0.21) &  & (0.25) &  \\ 
  $\;\;\;\;$ Outdegree popularity &  &  & 0.95 &  \\ 
   &  &  & (0.09) &  \\ 
  $\;\;\;\;$ Twopaths &  &  & -0.04 &  \\ 
   &  &  & (0.02) &  \\ 
  $\;\;\;\;$ GWIdegree (2.0) &  &  & 3.42 &  \\ 
   &  &  & (1.47) &  \\ 
  $\;\;\;\;$ GWESP (1.0) &  &  & 0.58 &  \\ 
   &  &  & (0.16) &  \\ 
  $\;\;\;\;$ GWOdegree (0.5) &  &  & 8.42 &  \\ 
   &  &  & (2.11) &  \\ 
   \hline\hline
\end{tabular}
\label{tab:regTable}
\end{table}
\FloatBarrier

There are also a few differences between the parameter estimates that result from the ERGM and AME. Using the AME we find evidence that \texttt{Preference dissimilarity} is associated with a reduced probability of collaboration between a pair of actors, which is in line with the theoretical expectations of Ingold \& Fischer.\footnote{See the Appendix for details.} Additionally, the AME results differ from ERGM for the nodal effects related to whether a receiver of a collaboration is a government actor, \texttt{Alter=Government actor}, and whether the sender is an environmental NGO, \texttt{Ego=Environmental NGO}.

\subsection*{Tie Formation Prediction}

To test which model more accurately captures the data generating process for this network, we utilize a cross-validation procedure to assess the out-of-sample performance for each of the models presented in Table~\ref{tab:regTable}. Our cross-validation approach proceeds as follows:

\begin{itemize}
	\item Randomly divide the $n \times (n-1)$ data points into $S$ sets of roughly equal size, letting $s_{ij}$ be the set to which pair $\{ij\}$ is assigned.
	\item For each $s \in \{1, \ldots, S\}$:
	\begin{itemize}
		\item Obtain estimates of the model parameters conditional on $\{y_{ij} : s_{ij} \neq s\}$, the data on pairs not in set $s$.
		\item For pairs $\{kl\}$ in set $s$, let $\hat y_{kl} = E[y_{kl} | \{y_{ij} : s_{ij} \neq s\}]$, the predicted value of $y_{kl}$ obtained using data not in set $s$.
	\end{itemize}
\end{itemize}

\begin{figure}[ht]
	\centering
	\caption{Assessments of out-of-sample predictive performance using ROC curves, separation plots, and PR curves. AUC statistics are also provided.}
	\begin{tabular}{cc}
	\includegraphics[width=.5\textwidth]{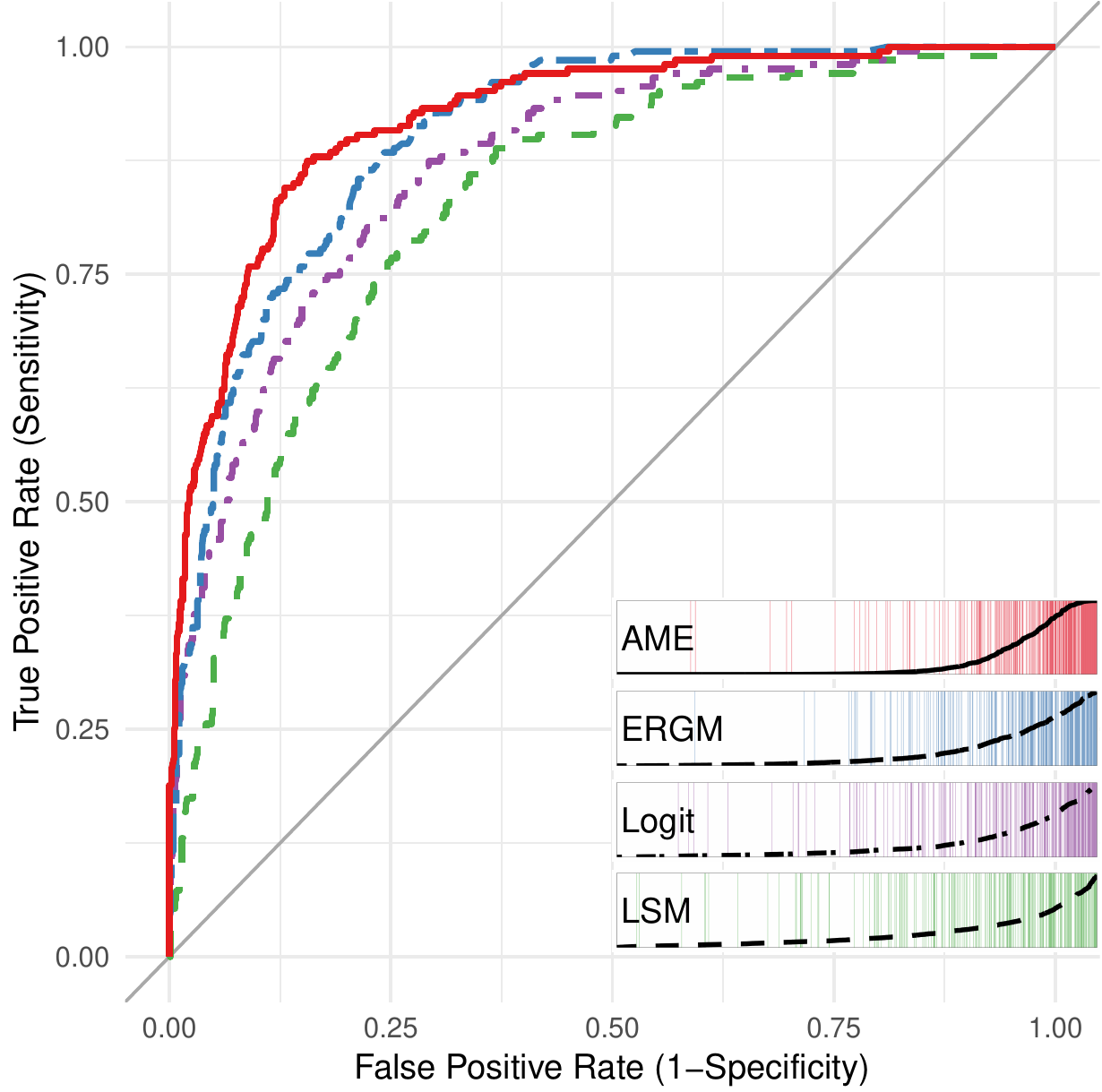} & 
	\includegraphics[width=.5\textwidth]{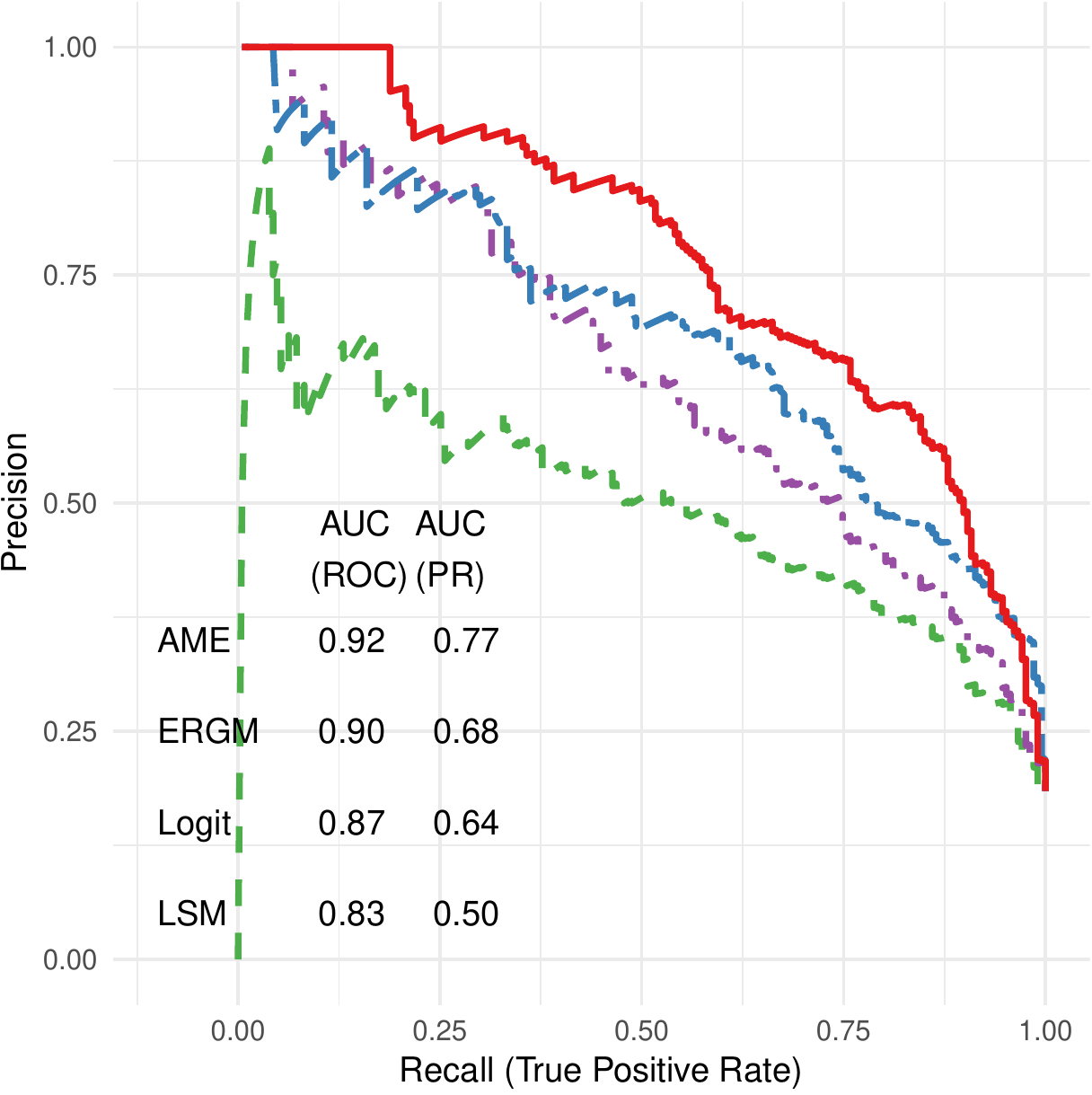}	
	\end{tabular}
	\label{fig:roc}
\end{figure}

The procedure summarized in the steps above generates a sociomatrix of out-of-sample predictions of the observed data. Each entry $\hat y_{ij}$ is a predicted value obtained from using a subset of the data that does not include $y_{ij}$. In this application we set $S$ to 45 which corresponds to randomly excluding approximately 2\% of the data from the estimation.\footnote{Such a low number of observations were excluded in every sample (denoted a fold) because excluding any more observations would cause the ERGM specification to result in a degenerate model that empirically can not be fit. This is an example of the computational difficulties associated with ERGMs.} Using the set of out-of-sample predictions we generate from the cross-validation procedure, we provide a series of tests to assess model fit. The left-most plot in Figure~\ref{fig:roc} compares the four approaches in terms of their ability to predict the out-of-sample occurrence of collaboration based on Receiver Operating Characteristic (ROC) curves. ROC curves provide a comparison of the trade-off between the True Positive Rate (TPR), sensitivity, False Positive Rate (FPR), 1-specificity, for each model. Models that have a better fit according to this test should have curves that follow the left-hand border and then the top border of the ROC space. On this diagnostic, the AME model performs best closely followed by ERGM. The Logit and LSM approach lag notably behind the other specifications. 

A more intuitive visualization of the differences between these modeling approaches can be gleaned through examining the separation plots included on the right-bottom edge of the ROC plot. This visualization tool plots each of the observations, in this case actor pairs, in the dataset according to their predicted value from left (low values) to right (high values). Models with a good fit should have all network links, here these are colored by the modeling approach, towards the right of the plot. Using this type of visualization emphasizes that the AME and ERGM models perform better than the alternatives.

The last diagnostic we highlight to assess predictive performance are precision-recall (PR) curves. In both ROC and PR space we utilize the TPR, also referred to as recall--though in the former it is plotted on the y-axis and the latter the x-axis. The difference, however, is that in ROC space we utilize the FPR, while in PR space we use precision. FPR measures the fraction of negative examples that are misclassified as positive, while precision measures the fraction of examples classified as positive that are truly positive. PR curves are useful in situations where correctly predicting events is more interesting than simply predicting non-events (\citealt{davis:goadrich:2006}). This is especially relevant in the context of studying many relational datasets in political science such as conflict, because events in such data are extremely sparse and it is relatively easy to correctly predict non-events. 

In the case of our application dataset, the vast majority of dyads, 80\%, do not have a network linkage, which points to the relevance of assessing performance using the PR curves as we do in the right-most plot of Figure~\ref{fig:roc}. We can see that the relative-ordering of the models remains similar but the differences in how well they perform become much more stark. Here we find that the AME approach performs notably better in actually predicting network linkages than each of the alternatives. Area under the curve (AUC) statistics are provided in Figure~\ref{fig:roc} and these also highlight AME's superior out-of-sample performance.\footnote{We also test AME against the multiple regression quadratic assignment procedure (MRQAP). This approach also perform notably worse than AME in terms of predicting tie formation. Results are available upon request.}

\FloatBarrier

\subsection*{Capturing Network Attributes}

\begin{figure}[ht]
	\centering
	\caption{Network goodness of fit summary using \pkg{amen}.}
	\includegraphics[width=1\textwidth]{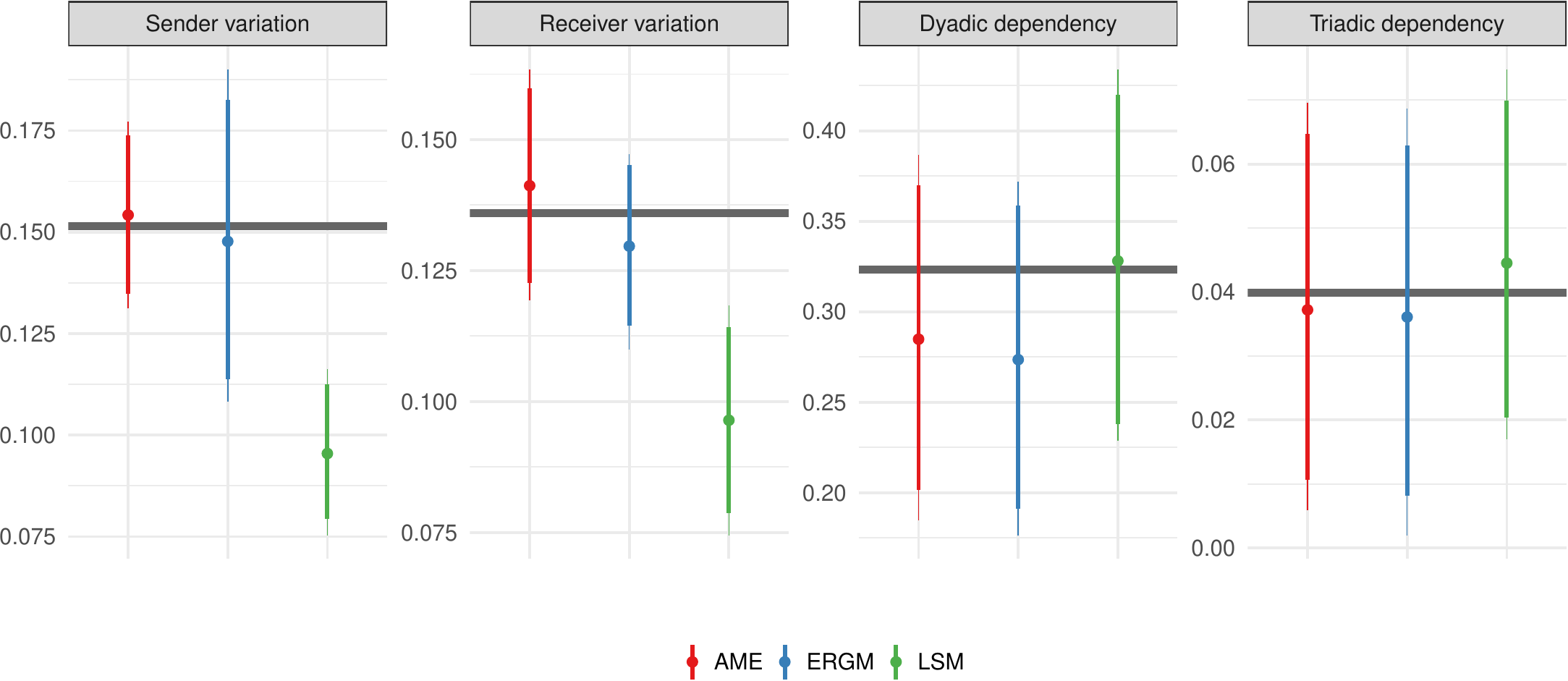}
	\label{fig:ergmAmePerf}
\end{figure}

We also assess which of these models best captures the network features of the dependent variable.\footnote{We restrict our focus to the three approaches--LSM, ERGM, and AME--that explicitly
seek to model network interdependencies.} To do this, we compare the observed network with a set of networks simulated from the estimated models.\footnote{In the Appendix, we compare the ability of these models to capture network attribute across a wider array of statistics (e.g., dyad-wise shared partners, incoming k-star, etc.), and the results are consistent with what we present below.} We simulate 1,000 networks from the three models and compare how well they align with the observed network in terms of four network statistics: (1) the empirical standard deviation of the row means (i.e., heterogeneity of nodes in terms of the ties they send); (2) the empirical standard deviation of the column means (i.e., heterogeneity of nodes in terms of the ties they receive); (3) the empirical within-dyad correlation (i.e., measure of reciprocity in the network); and (4) a normalized measure of triadic dependence. A comparison of the LSM, ERGM, and AME models among these four statistics is shown in Figure~\ref{fig:ergmAmePerf}.

\FloatBarrier

Here it becomes quickly apparent that the LSM model fails to capture how active and popular actors are in the Swiss climate change mitigation network.\footnote{Further even after incorporating random sender and receiver effects into the LSM framework this problem is not completely resolved, see the Appendix for details.} The AME and ERGM specifications again both tend to do equally well. If when running this diagnostic, we found that the AME model did not adequately represent the observed network this would indicate that we might want to increase $K$ to better account for network interdependencies. No changes to the model specification as described by the exogenous covariates a researcher has chosen would be necessary. If the ERGM results did not align with the diagnostic presented in Figure~\ref{fig:ergmAmePerf}, then this would indicate that an incorrect set of endogenous dependencies have been specified. 

\section*{\textbf{Conclusion}}

The AME approach to estimation and inference in network data provides a number of benefits over extant alternatives in political science. Specifically, it provides a modeling framework for dyadic data that is based on familiar statistical tools such as linear regression, GLM, random effects, and factor models.\footnote{A number of related approaches have been developed that also stem from latent variable models: \citet{sewell:chen:2015,gollini:murphy:2016,durante:etal:2017,kao:etal:2018}. Each of these approaches differ in how they construct the latent variable term to account for third-order dependencies, but they each are based off of a similar framework as the model we present here. We hope that this paper motivates further interest in exploring the utility of latent variable models to studying networks in political science.} Further we have shown that alternatives such as the LSM complicate parameter interpretation due to the construction of the latent variable term. The benefit of AME is that its focus intersects with the interest of most IR scholars, which is primarily on the effects of exogenous covariates. For researchers in the social sciences this is of primary interest, as many studies that employ relational data still have conceptualizations that are monadic or dyadic in nature.

ERGMs are best suited for cases in which scholars are interested in studying the role that particular types of node- and dyad-based network configurations play in generating the network. Though valuable this is often orthogonal to the interest of most researchers who are focused on studying the affect of a particular exogenous variable, such as democracy, on a dyadic variable like conflict while simply accounting for network dependencies. Additionally, through the application dataset utilized herein we show that the AME approach outperforms both ERGM and LSM in out-of-sample prediction, and also is better able to capture network dependencies than the LSM.

More broadly, relational data structures are composed of actors that are part of a system.\footnote{Additionally, in most political science applications, we are interested in how actors behave towards each other over time. Accounting for repeated interaction within AME can be done by including time-dependent regression terms such as lags of the dependent variable or simply time-varying regression parameters.} It is unlikely that this system can be viewed simply as a collection of isolated actors or pairs of actors. The assumption  that dependencies between observations occur can at the very least be examined. Failure to take into account interdependencies leads to biased parameter estimates and poor fitting models. By using standard diagnostics such as shown in Figure~\ref{fig:ergmAmePerf}, one can easily assess whether an assumption of independence is reasonable. We stress this point because a common misunderstanding that seems to have emerged within the social science literature relying on dyadic data is that a network based approach is only necessary if one has theoretical explanations that extend beyond the dyadic. This is not at all the case and findings that continue to employ a dyadic design may misrepresent the effects of the very variables that they are interested in. The AME approach that we have detailed here provides a statistically familiar way for scholars to account for unobserved network structures in relational data. 

\newpage
\clearpage

\renewcommand{\thefigure}{A\arabic{figure}}
\setcounter{figure}{0}
\renewcommand{\thetable}{A.\arabic{table}}
\setcounter{table}{0}
\renewcommand{\thesection}{A.\arabic{section}}
\setcounter{section}{0}	

\section*{Appendix}
\clearpage

\subsection*{Additive and Multiplicative Effects Gibbs Sampler}

To estimate, the effects of our exogenous variables and latent attributes we utilize a Bayesian probit model in which we sample from the posterior distribution of the full conditionals until convergence. Specifically, given observed data $\textbf{Y}$ and $\textbf{X}$ -- where $\textbf{X}$ is a design array that includes our sender, receiver, and dyadic covariates -- we estimate our network of binary ties using a probit framework where: $y_{ij,t} = 1(\theta_{ij,t}>0)$ and $\theta_{ij,t} = \bm\beta^{\top}\mathbf{X}_{ij,t} + a_{i} + b_{j} + \textbf{u}_{i}^{\top} \textbf{D} \textbf{v}_{j} + \epsilon_{ij}$. The derivation of the full conditionals is described in detail in \citet{hoff:2005} and \citet{hoff:2008}, thus here we only outline the Markov chain Monte Carlo (MCMC) algorithm for the AME model that we utilize in this paper.

\begin{itemize}
 \item Given initial values of $\{\bm\beta, \textbf{a}, \textbf{b}, \textbf{U}, \textbf{V}, \Sigma_{ab}, \rho, \text{ and } \sigma_{\epsilon}^{2}\}$, the algorithm proceeds as follows:
 \begin{itemize}
 	\item sample $\bm\theta \; | \;  \bm\beta, \textbf{X}, \bm\theta, \textbf{a}, \textbf{b}, \textbf{U}, \textbf{V}, \Sigma_{ab}, \rho, \text{ and } \sigma_{\epsilon}^{2}$ (Normal)
 	\item sample $\bm\beta \; | \;  \textbf{X}, \bm\theta, \textbf{a}, \textbf{b}, \textbf{U}, \textbf{V}, \Sigma_{ab}, \rho, \text{ and } \sigma_{\epsilon}^{2}$ (Normal)
 	\item sample $\textbf{a}, \textbf{b} \; | \; \bm\beta, \textbf{X}, \bm\theta, \textbf{U}, \textbf{V}, \Sigma_{ab}, \rho, \text{ and } \sigma_{\epsilon}^{2}$ (Normal)
	\item sample $\Sigma_{ab} \; | \; \bm\beta, \textbf{X}, \bm\theta, \textbf{a}, \textbf{b}, \textbf{U}, \textbf{V}, \rho, \text{ and } \sigma_{\epsilon}^{2}$ (Inverse-Wishart)
 	\item update $\rho$ using a Metropolis-Hastings step with proposal $p^{*} | p  \sim$ truncated normal$_{[-1,1]}(\rho, \sigma_{\epsilon}^{2})$
 	\item sample $\sigma_{\epsilon}^{2} \; | \; \bm\beta, \textbf{X}, \bm\theta, \textbf{a}, \textbf{b}, \textbf{U}, \textbf{V}, \Sigma_{ab}, \text{ and } \rho$ (Inverse-Gamma)
 	\item For each $k \in K$:
 	\begin{itemize}
 		\item Sample $\textbf{U}_{[,k]} \; | \; \bm\beta, \textbf{X}, \bm\theta, \textbf{a}, \textbf{b}, \textbf{U}_{[,-k]}, \textbf{V}, \Sigma_{ab}, \rho, \text{ and } \sigma_{\epsilon}^{2}$ (Normal)
 		\item Sample $\textbf{V}_{[,k]} \; | \; \bm\beta, \textbf{X}, \bm\theta, \textbf{a}, \textbf{b}, \textbf{U}, \textbf{V}_{[,-k]}, \Sigma_{ab}, \rho, \text{ and } \sigma_{\epsilon}^{2}$ (Normal)
 		\item Sample $\textbf{D}_{[k,k]}  \; | \; \bm\beta, \textbf{X}, \bm\theta, \textbf{a}, \textbf{b}, \textbf{U}, \textbf{V}, \Sigma_{ab}, \rho, \text{ and } \sigma_{\epsilon}^{2}$ (Normal)\footnote{Subsequent to estimation, \textbf{D} matrix is absorbed into the calculation for $\textbf{V}$ as we iterate through $K$. }
 	\end{itemize}
 \end{itemize}
\end{itemize}

\clearpage
\subsection*{Ingold \& Fischer Model Specification and Expected Effects}

\newcolumntype{L}{>{\arraybackslash}m{9cm}}
\begin{table}[ht]
\centering
\begingroup\scriptsize
\begin{tabular}{lLc}
\footnotesize{\textbf{Variable}} & \footnotesize{\textbf{Description}} & \footnotesize{\textbf{Expected Effect}} \\ \hline\hline
	\multicolumn{3}{l}{\textbf{Conflicting policy preferences}} \\
	\quad Business v. NGO & Binary, dyadic covariate that equals one when one actor is from the business sector and the other an NGO. & $-$ \\
	\quad Opposition/alliance & Binary, dyadic covariate that equals one when $i$, sender, perceives $j$, receiver, as having similar policy objectives regarding climate change.  & $+$ \\
	\quad Preference dissimilarity & Transformation of four core beliefs into a Manhattan distance matrix, smaller the distance the closer the beliefs of $i$ and $j$. & $-$ \\
	\multicolumn{3}{l}{\textbf{Transaction costs}} \\
	\quad Joint forum participation & Binary, dyadic covariate that equals one when $i$ and $j$ belong to the same policy forum. & $+$ \\
	\multicolumn{3}{l}{\textbf{Influence}} \\
	\quad Influence attribution & Binary, dyadic covariate that equals one when $i$ considers $j$ to be influential. & $+$ \\
	\quad Alter's influence in-degree & Number of actors that mention $i$ as being influential, this is a measure of reputational power. & $+$ \\
	\quad Influence absolute diff. & Absolute difference in reputational power between $i$ and $j$. & $-$ \\
	\quad Alter = Government Actor & Binary, nodal covariate that equals one when $j$ is a state actor. & $+$ \\
	\multicolumn{3}{l}{\textbf{Functional requirements}} \\
	\quad Ego = Environment NGO & Binary, nodal covariate that equals one when $i$ is an NGO. & $+$ \\
	\quad Same actor type & Binary, dyadic covariate that equals when $i$ and $j$ are the same actor type. & $+$ \\
	\multicolumn{3}{l}{\textbf{Endogenous dependencies: ERGM Specific Parameters}} \\
	\quad Mutuality & Captures concept of reciprocity, if $i$ indicates they collaborated with $j$ then $j$ likely collaborates with $i$. & $+$\\
	\quad Outdegree popularity & Captures idea that actors sending more ties will be more popular targets themselves for collaboration.  & $+$ \\
	\quad Twopaths & Counts the number of two-paths in the network, two-path is an instance where $i$ is connected to $j$, $j$ to $k$, but $i$ is not connected to $k$. & $-$ \\
	\quad GWIdegree (2.0) & Takes into account how many ties a node sends in the network, used to capture network structures that result from some highly active nodes.  & $+$ \\
	\quad GWESP (1.0) & Counts the number of shared partners for each pair and sums across.  & $+$ \\
	\quad GWOdegree (0.5) & Takes into account how many ties a node receives in the network, used to capture networks structures that result from some highly popular nodes.  & $+$ \\
\hline\hline
\end{tabular}
\endgroup
\caption{Summary of variables to be included in model specification.}
\label{tab:theorySpec}
\end{table}
\FloatBarrier

\clearpage
\subsection*{AME Model Convergence}
\label{sec:ameConvAppendix}

Trace plot for AME model presented in paper.

\begin{figure}[ht]
	\centering
	\begin{tabular}{cc}
	\includegraphics[width=.45\textwidth]{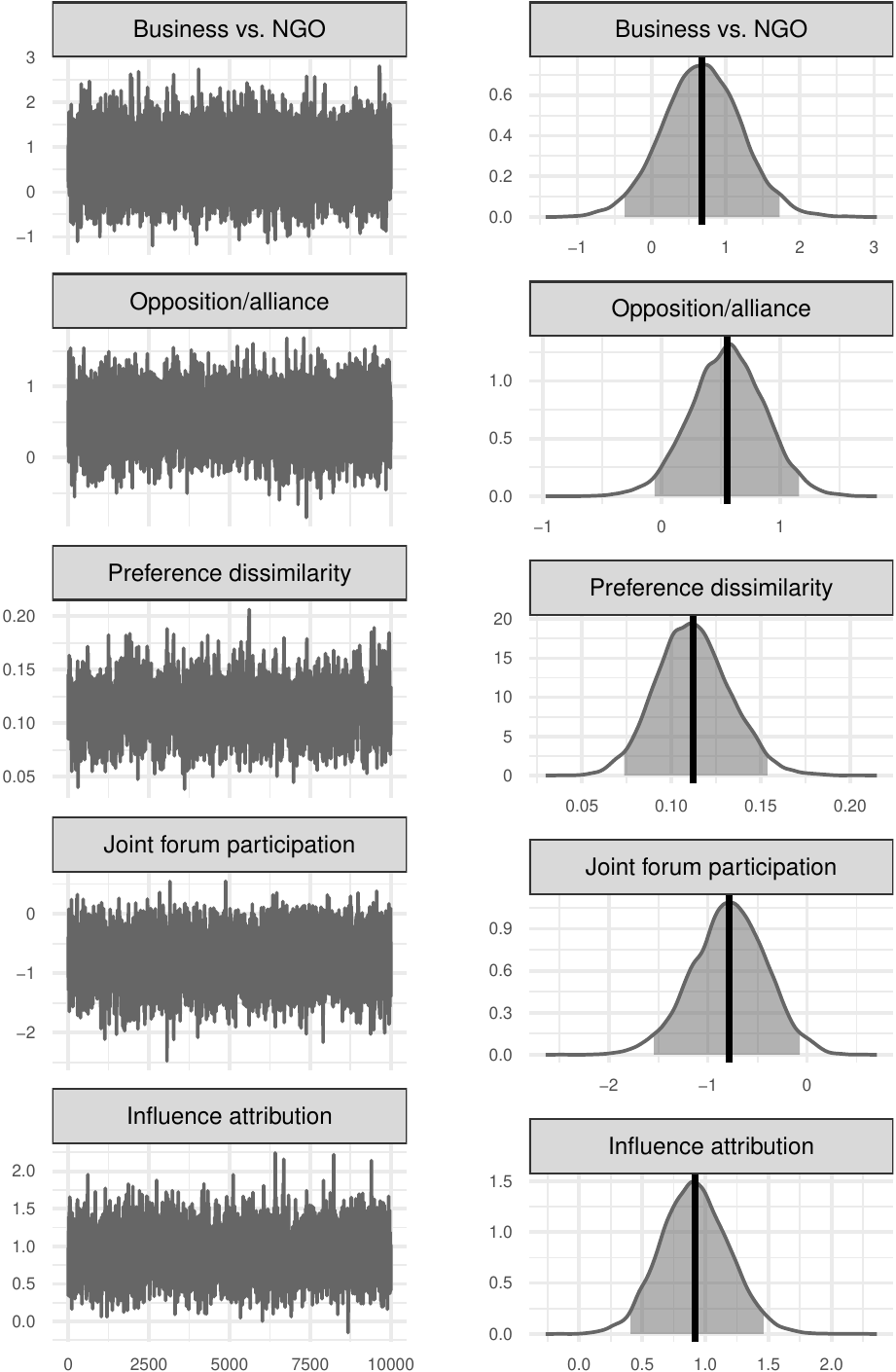} &
	\includegraphics[width=.45\textwidth]{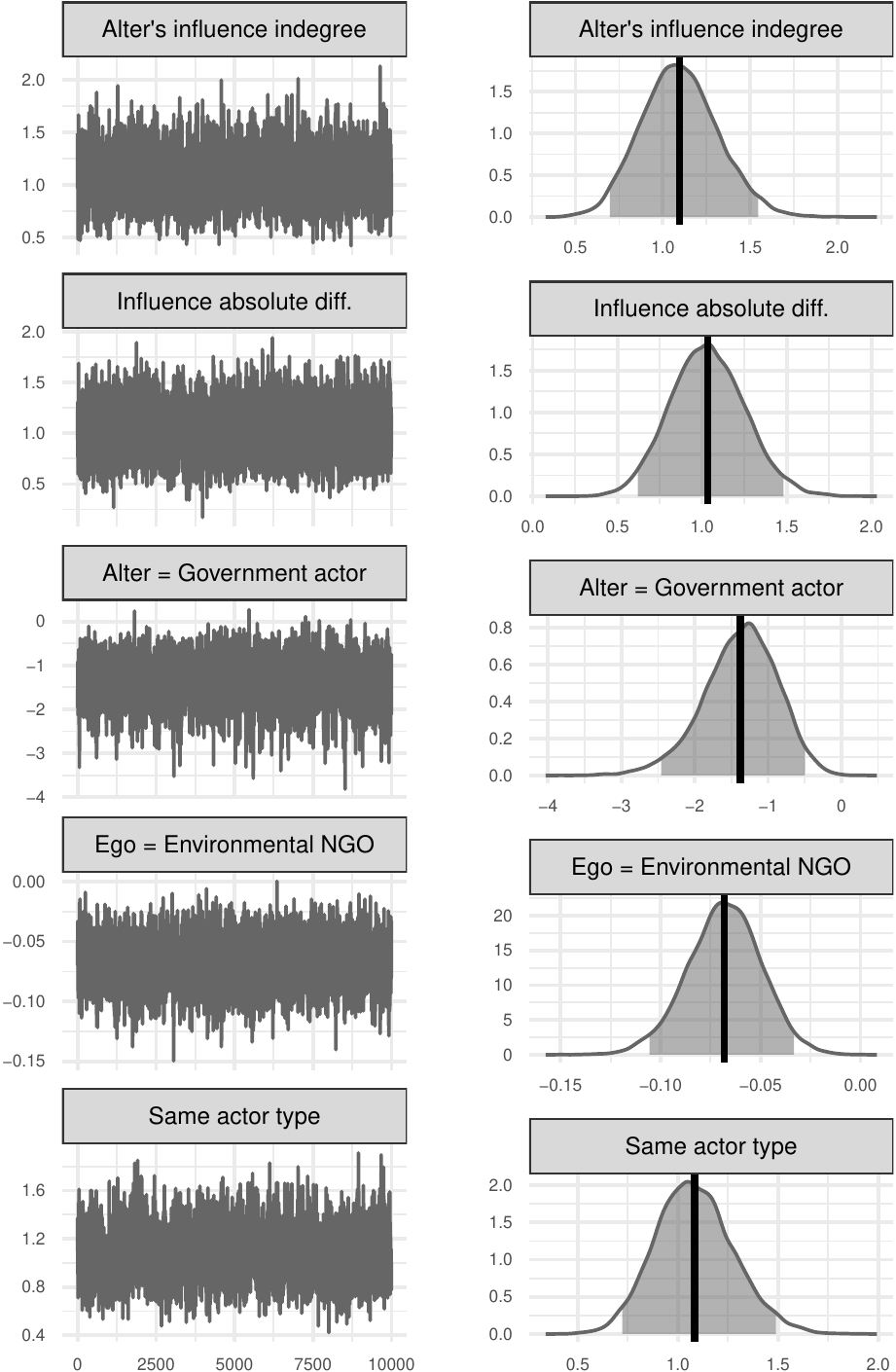}
	\end{tabular}
	\caption{Trace plot for AME model presented in paper. In this model, we utilize the SRM to account for first and second-order dependence. To account for third order dependencies we use the latent factor approach with $K=2$.}
	\label{fig:ameConv}
\end{figure}
\FloatBarrier
\newpage

\subsection*{Multiplicative Effects Visualization}

\begin{figure}[ht]
\centering
	\includegraphics[width=.5\textwidth]{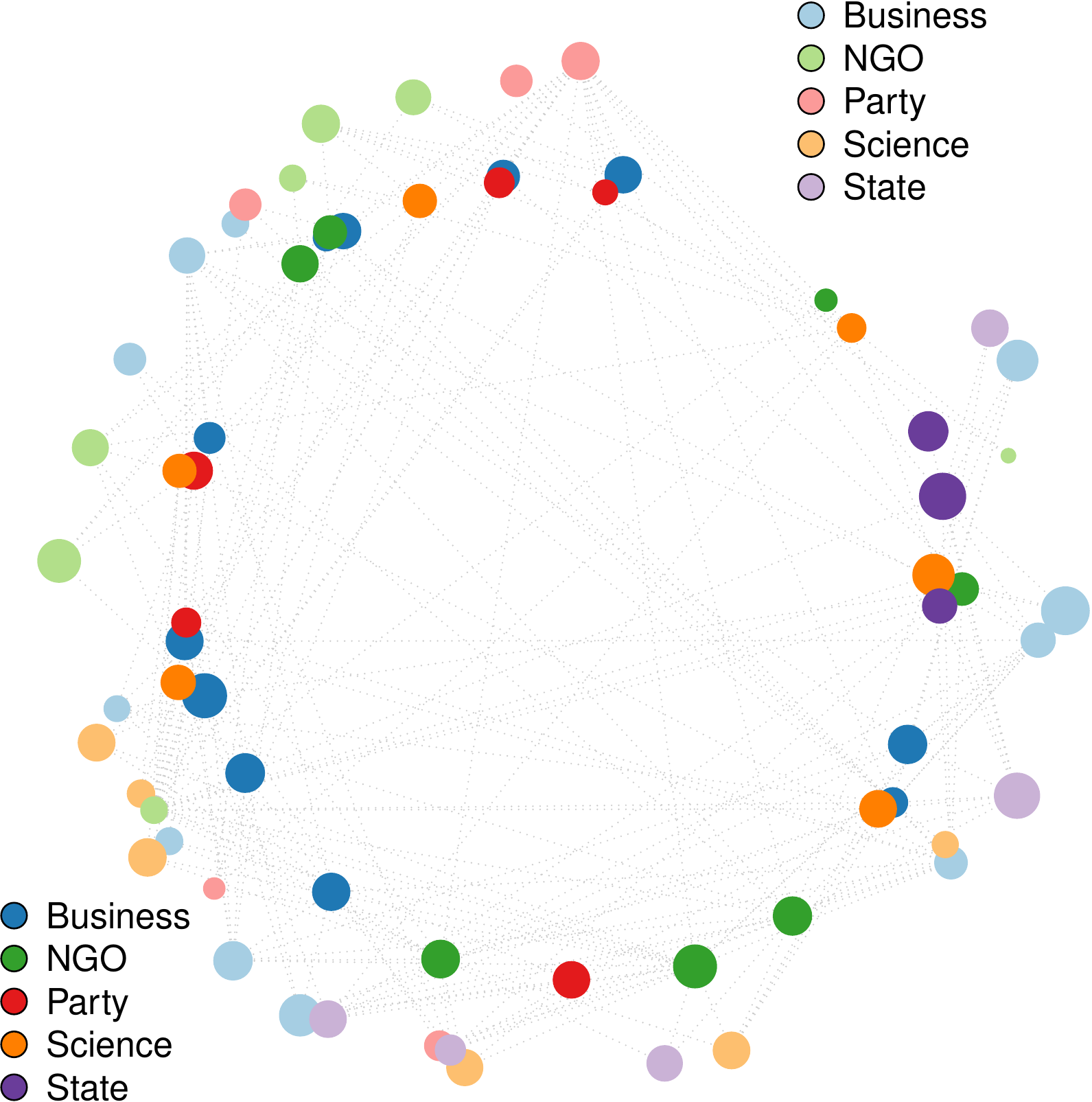}
	\caption{Circle plot of estimated latent factors.}
	\label{fig:uv}
\end{figure}

When it comes to estimating higher-order effects, ERGM is able to provide explicit estimates of a variety of higher-order parameters, however, this comes with the caveat that these are the ``right'' set of endogenous dependencies. The AME approach, as shown in Equation 4 of the manuscript, estimates network dependencies by examining patterns left over after taking into account the observed covariates. For the sake of space, we focus on examining the third-order dependencies left over after accounting for the observed covariates and network covariance structure modeled by the SRM. A visualization of remaining third-order dependencies is shown in Figure~\ref{fig:uv}.

\FloatBarrier

In Figure~\ref{fig:uv}, the directions of $\hat{u}_{i}$'s and $\hat{v}_{i}$'s are noted in lighter and darker shades, respectively, of an actor's type.\footnote{For example, actors from industry and business are assigned a color of blue and the direction of $\hat{u}_{i}$ for these actors is shown in light blue and $\hat{v}_{i}$ in dark blue} The size of actors is a function of the magnitude of the vectors, and dashed lines between actors indicate greater than expected levels of collaboration based on the regression term and additive effects. In the case of the application dataset that we are using here organization names have been anonymized and no additional covariate information is available. However, if we were to observe nodes sharing certain attributes clustering together in this circle plot that would mean such an attribute could be an important factor in helping us to understand collaborations among actors in this network. Given how actors of different types are distributed in almost a random fashion in this plot, we can at least be sure that it is unlikely other third-order patterns can be picked up by that factor.

\clearpage
\subsection*{Other Network Goodness of Fit Tests}
\label{sec:otherNetGof}

Below we show a standard set of statistics upon which comparisons are usually conducted:\footnote{See \citet{morris:etal:2008} for details on each of these parameters. If one was to examine goodness of fit in the \pkg{ergm} package these parameters would be calculated by default.}

\newcolumntype{L}{>{\arraybackslash}m{9cm}}
\begin{table}[ht]
\centering
\begingroup\scriptsize
\begin{tabular}{lL}
\footnotesize{\textbf{Variable}} & \footnotesize{\textbf{Description}} \\ \hline\hline
	Dyad-wise shared partners & Number of dyads in the network with exactly $i$ shared partners. \\
	Edge-wise shared partners & Similar to above except this counts the number of dyads with the same number of edges. \\
	Geodesic distances & The proportion of pairs of nodes whose shortest connecting path is of length $k$, for $k=1,2,\ldots$. Also, pairs of nodes that are not connected are classified as $k=\infty$. \\
	Incoming k-star & Propensities for individuals to have connections with multiple network partners. \\
	Indegree & Proportion of nodes with the same value of the attribute as the receiving node. \\
	Outdegree & Proportion of nodes with the same value of the attribute as the sending node. \\
\hline\hline
\end{tabular}
\endgroup
\caption{Description of a set of standard statistics used to assess whether a model captures network dependencies. }
\label{tab:netStat}
\end{table}
\FloatBarrier

\begin{figure}[ht]
	\centering
	\includegraphics[width=1\textwidth]{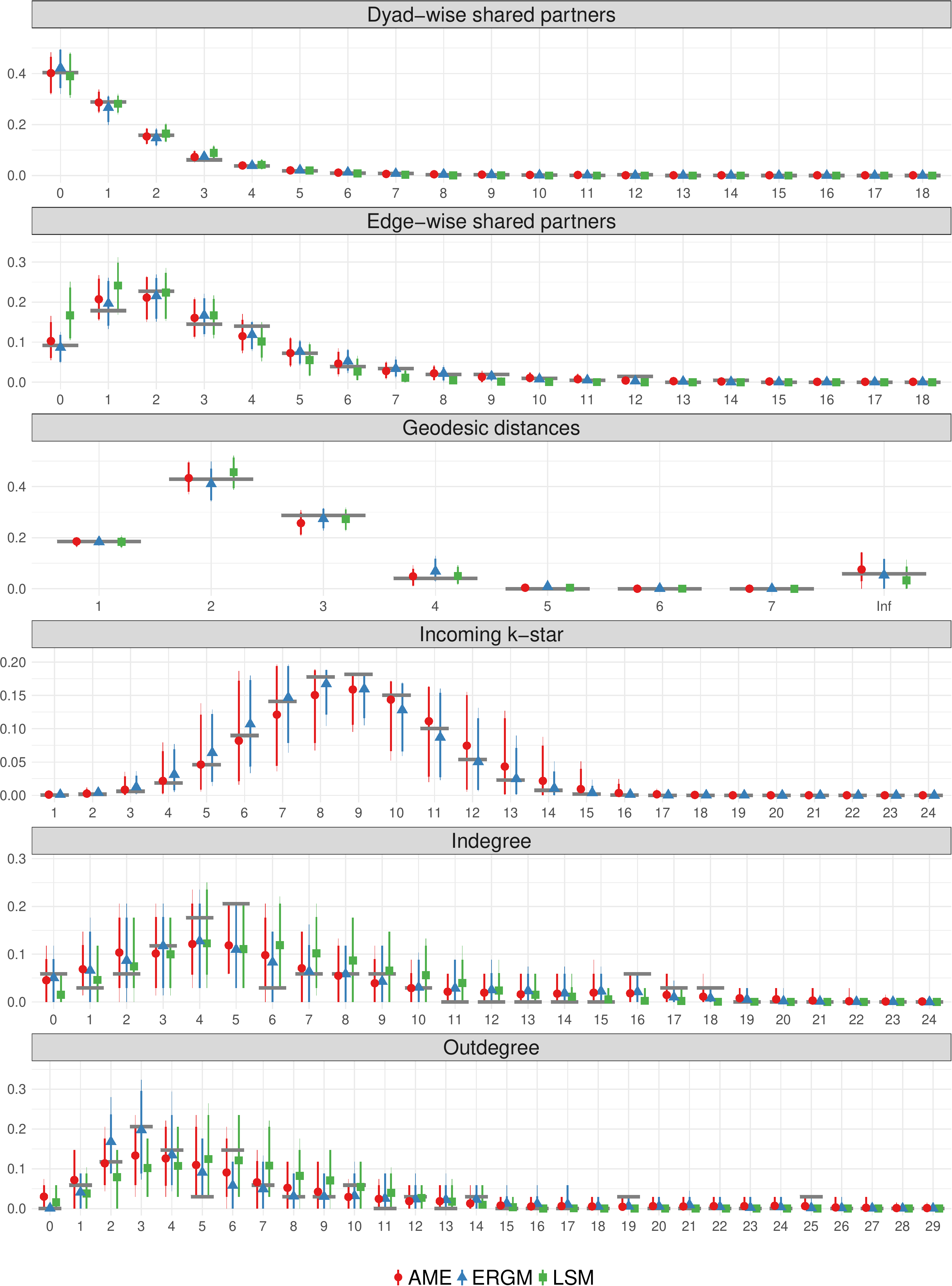}
	\caption{Goodness of fit statistics to assess how well the LSM, ERGM, and AME approaches account for network dependencies. Grey bars indicate true values.}
	\label{fig:gofAll}
\end{figure}

We simulate 1,000 networks from the LSM, ERGM, and AME model and compare how well they align with the observed network in terms of the statistics described in Table~\ref{tab:netStat}. The results are shown in Figure~\ref{fig:gofAll}. Values for the observed network are indicated by a gray bar and average values from the simulated networks for the AME, ERGM, and LSM are represented by a diamond, triangle, and square, respectively. The densely shaded interval around each point represents the 95\% interval from the simulations and the taller, less dense the 90\% interval.\footnote{Calculation for the incoming k-star statistic is not currently supported by the \pkg{latentnet} package.} Looking across the panels in Figure~\ref{fig:gofAll} it is clear that there is little difference between the ERGM and AME models in terms of how well they capture network dependencies. The LSM model, however, does perform somewhat worse in comparison here as well. Particularly, when it comes to assessing the number of edge-wise shared partners and in terms of capturing the indegree and outdegree distributions of the collaboration network.

\FloatBarrier

\clearpage
\subsection*{Comparison with other AME Parameterizations}
\label{sec:ameVsAmeAppendix}

Here we provide a comparison of the AME model we present in the paper that uses $K=2$ for multiplicative effects and show how results change when we use $K=\{1,3,4\}$. Trace plots for $K=\{1,3,4\}$ are available upon request.

\begin{table}[ht]
\centering
\begingroup\footnotesize
\begin{tabular}{lcccc}
   & AME (k=1) & AME (k=2) & AME (k=3) & AME (k=4) \\
  \hline
\hline
Intercept/Edges & -3.08 & -3.40 & -3.74 & -3.92 \\ 
   & [-3.91; -2.30] & [-4.40; -2.51] & [-4.80; -2.75] & [-5.13; -2.87] \\ 
  \textbf{Conflicting policy preferences} &  &  &  &  \\ 
  $\;\;\;\;$ Business vs. NGO & -1.28 & -1.38 & -1.51 & -1.50 \\ 
   & [-2.20; -0.46] & [-2.47; -0.49] & [-2.64; -0.53] & [-2.69; -0.52] \\ 
  $\;\;\;\;$ Opposition/alliance & 0.95 & 1.08 & 1.19 & 1.27 \\ 
   & [0.65; 1.28] & [0.72; 1.49] & [0.80; 1.63] & [0.84; 1.76] \\ 
  $\;\;\;\;$ Preference dissimilarity & -0.65 & -0.79 & -0.92 & -0.96 \\ 
   & [-1.29; -0.02] & [-1.55; -0.07] & [-1.76; -0.12] & [-1.80; -0.15] \\ 
  \textbf{Transaction costs} &  &  &  &  \\ 
  $\;\;\;\;$ Joint forum participation & 0.84 & 0.92 & 1.02 & 1.05 \\ 
   & [0.37; 1.31] & [0.40; 1.46] & [0.46; 1.62] & [0.43; 1.73] \\ 
  \textbf{Influence} &  &  &  &  \\ 
  $\;\;\;\;$ Influence attribution & 1.00 & 1.10 & 1.21 & 1.27 \\ 
   & [0.64; 1.39] & [0.70; 1.55] & [0.77; 1.70] & [0.80; 1.83] \\ 
  $\;\;\;\;$ Alter's influence indegree & 0.10 & 0.11 & 0.13 & 0.13 \\ 
   & [0.07; 0.14] & [0.07; 0.15] & [0.08; 0.17] & [0.09; 0.18] \\ 
  $\;\;\;\;$ Influence absolute diff. & -0.06 & -0.07 & -0.08 & -0.08 \\ 
   & [-0.10; -0.03] & [-0.11; -0.03] & [-0.13; -0.04] & [-0.12; -0.04] \\ 
  $\;\;\;\;$ Alter = Government actor & 0.52 & 0.56 & 0.66 & 0.63 \\ 
   & [-0.04; 1.07] & [-0.06; 1.16] & [-0.04; 1.46] & [-0.08; 1.36] \\ 
  \textbf{Functional requirements} &  &  &  &  \\ 
  $\;\;\;\;$ Ego = Environmental NGO & 0.61 & 0.68 & 0.77 & 0.79 \\ 
   & [-0.31; 1.56] & [-0.36; 1.73] & [-0.35; 1.93] & [-0.38; 2.04] \\ 
  $\;\;\;\;$ Same actor type & 0.97 & 1.03 & 1.14 & 1.18 \\ 
   & [0.60; 1.36] & [0.62; 1.48] & [0.66; 1.66] & [0.70; 1.70] \\ 
   \hline
\hline
\end{tabular}
\endgroup
\caption{95\% posterior credible intervals are provided in brackets.}
\label{tab:regTable_ame}
\end{table}

\begin{figure}[ht]
	\centering
	\begin{tabular}{cc}
	\includegraphics[width=.5\textwidth]{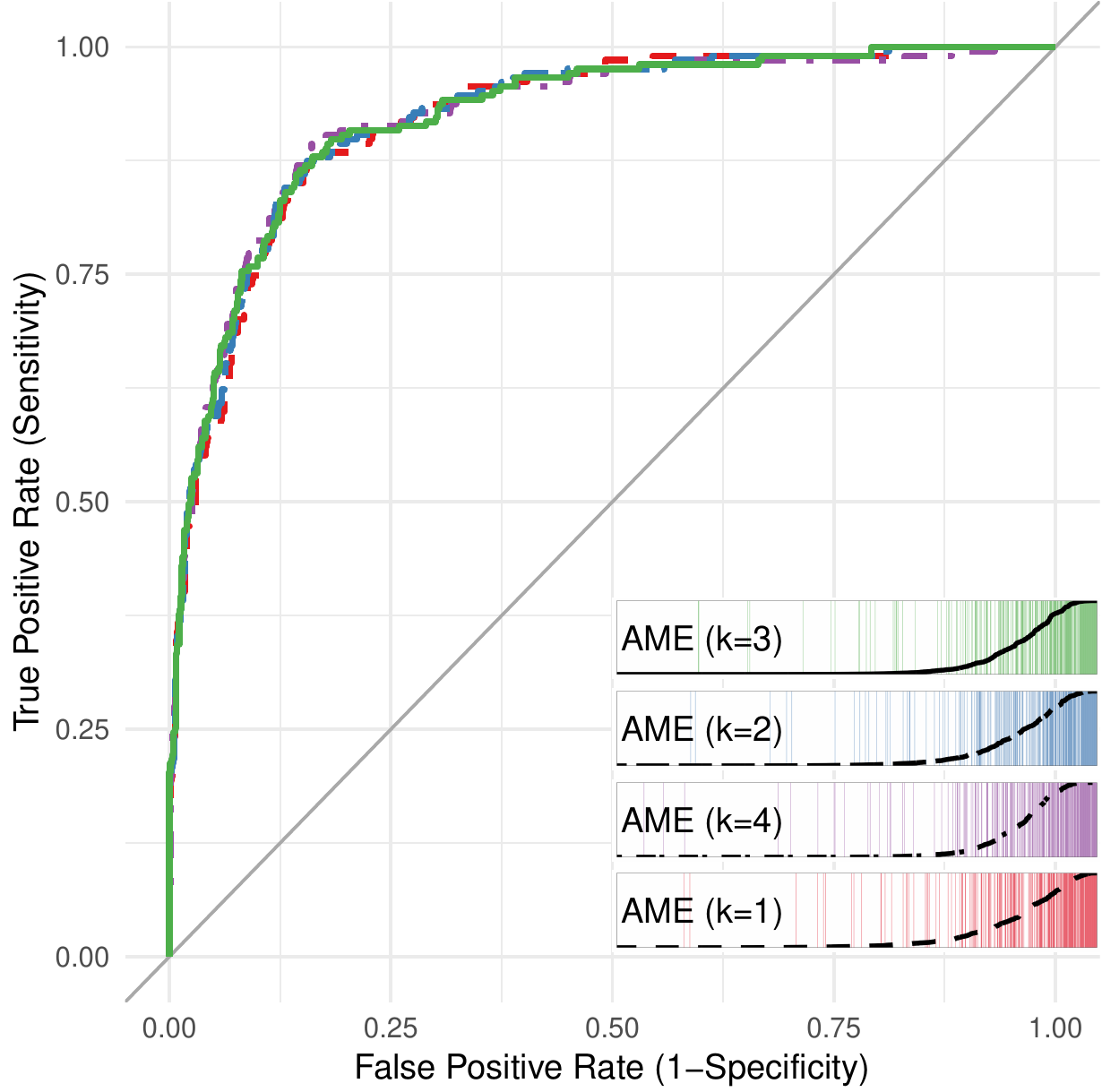} &
	\includegraphics[width=.5\textwidth]{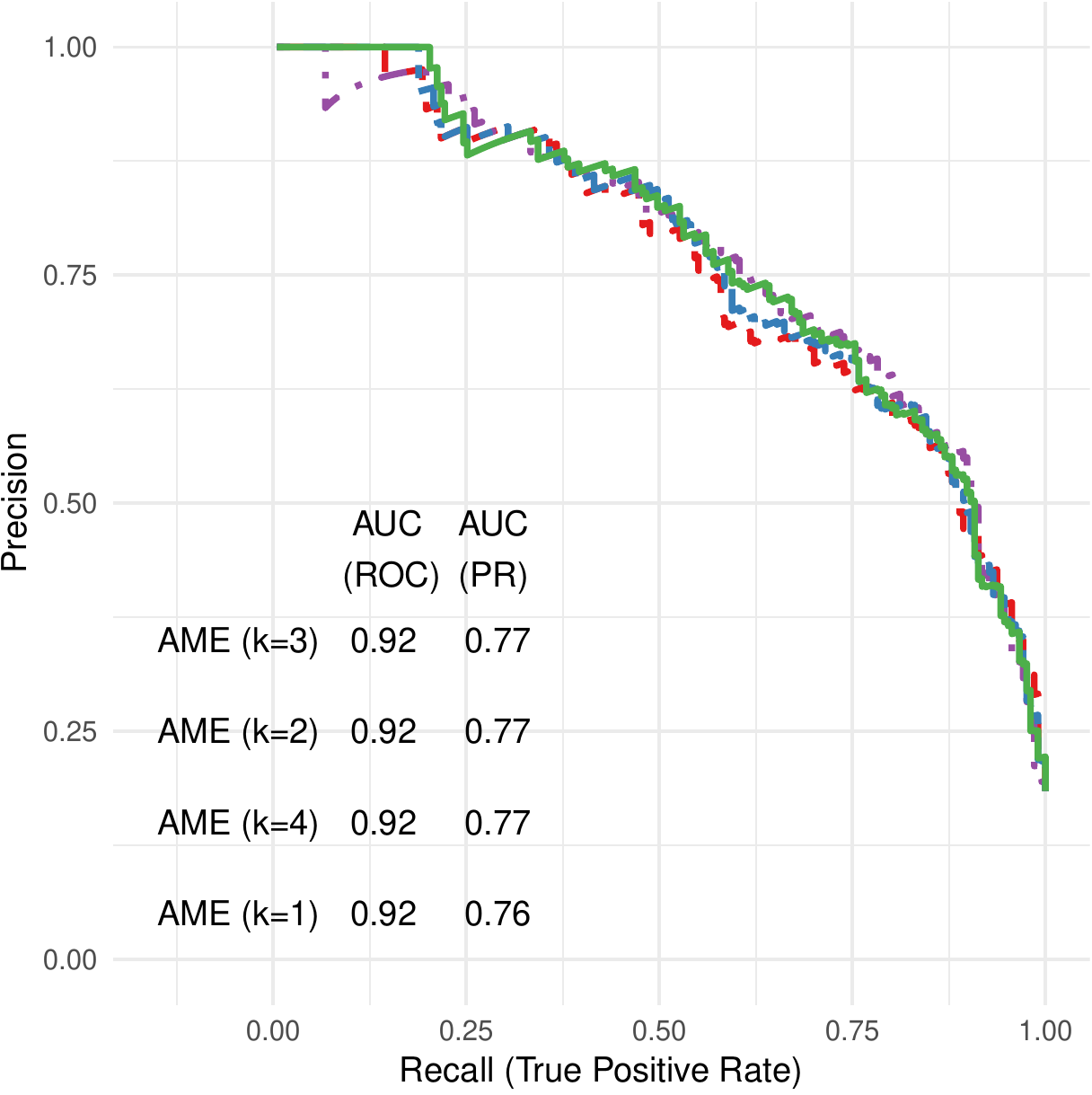}
	\end{tabular}
	\caption{Assessments of out-of-sample predictive performance using ROC curves, separation plots, and PR curves. AUC statistics are provided as well for both curves.}
	\label{fig:roc_ame}
\end{figure}

\begin{figure}[ht]
	\centering
	\includegraphics[width=1\textwidth]{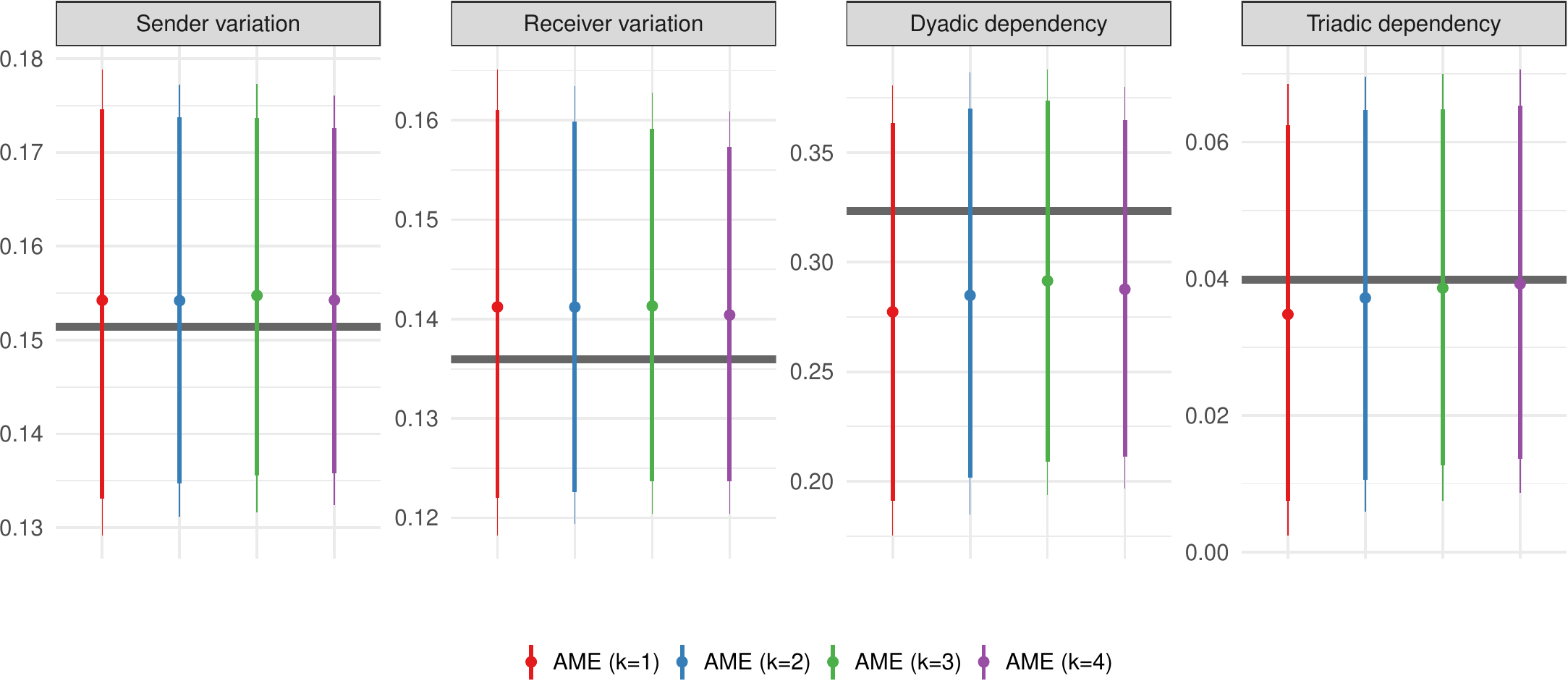}
	\caption{Network goodness of fit summary using \pkg{amen}.}
	\label{fig:netPerfCoef_ameSR}
\end{figure}

\clearpage
\subsection*{Comparison of \pkg{amen} \& \pkg{latentnet} $\sf{R}$ Packages}
\label{sec:ameVsLatentnetAppendix}

Here we provide a comparison of the AME model we present in the paper with a variety of parameterizations from the \pkg{latentnet} package. The number of dimensions in the latent space in each of these cases is set to 2. LSM (SR) represents a model in which random sender and receiver effects are included. 

\begin{table}[ht]
\centering
\begingroup\footnotesize
\begin{tabular}{lccc}
   & LSM & LSM (SR) & AME \\ 
  \hline
\hline
Intercept/Edges & 0.95 & 0.61 & -3.40 \\ 
   & [0.09; 1.85] & [-1.07; 2.34] & [-4.40; -2.51] \\ 
  \textbf{Conflicting policy preferences} &  &  &  \\ 
  $\;\;\;\;$ Business vs. NGO & -1.37 & -3.06 & -1.38 \\ 
   & [-2.39; -0.40] & [-4.72; -1.59] & [-2.47; -0.49] \\ 
  $\;\;\;\;$ Opposition/alliance & 0.00 & 0.30 & 1.08 \\ 
   & [-0.40; 0.40] & [-0.26; 0.86] & [0.72; 1.49] \\ 
  $\;\;\;\;$ Preference dissimilarity & -1.77 & -1.88 & -0.79 \\ 
   & [-2.64; -0.91] & [-3.06; -0.69] & [-1.55; -0.07] \\ 
  \textbf{Transaction costs} &  &  &  \\ 
  $\;\;\;\;$ Joint forum participation & 1.51 & 1.55 & 0.92 \\ 
   & [0.85; 2.17] & [0.66; 2.40] & [0.40; 1.46] \\ 
  \textbf{Influence} &  &  &  \\ 
  $\;\;\;\;$ Influence attribution & 0.08 & 0.30 & 1.10 \\ 
   & [-0.40; 0.54] & [-0.38; 0.96] & [0.70; 1.55] \\ 
  $\;\;\;\;$ Alter's influence indegree & 0.01 & 0.06 & 0.11 \\ 
   & [-0.03; 0.04] & [-0.03; 0.14] & [0.07; 0.15] \\ 
  $\;\;\;\;$ Influence absolute diff. & 0.04 & -0.08 & -0.07 \\ 
   & [-0.01; 0.09] & [-0.14; -0.02] & [-0.11; -0.03] \\ 
  $\;\;\;\;$ Alter = Government actor & -0.46 & -0.09 & 0.56 \\ 
   & [-1.08; 0.14] & [-1.85; 1.77] & [-0.06; 1.16] \\ 
  \textbf{Functional requirements} &  &  &  \\ 
  $\;\;\;\;$ Ego = Environmental NGO & -0.60 & -1.72 & 0.68 \\ 
   & [-1.30; 0.08] & [-3.76; 0.25] & [-0.36; 1.73] \\ 
  $\;\;\;\;$ Same actor type & 1.17 & 1.81 & 1.03 \\ 
   & [0.62; 1.72] & [1.09; 2.56] & [0.62; 1.48] \\ 
   \hline
\hline
\end{tabular}
\endgroup
\caption{95\% posterior credible intervals are provided in brackets.}
\label{tab:regTable_latSpace}
\end{table}

\begin{figure}[ht]
	\centering
	\begin{tabular}{cc}
	\includegraphics[width=.5\textwidth]{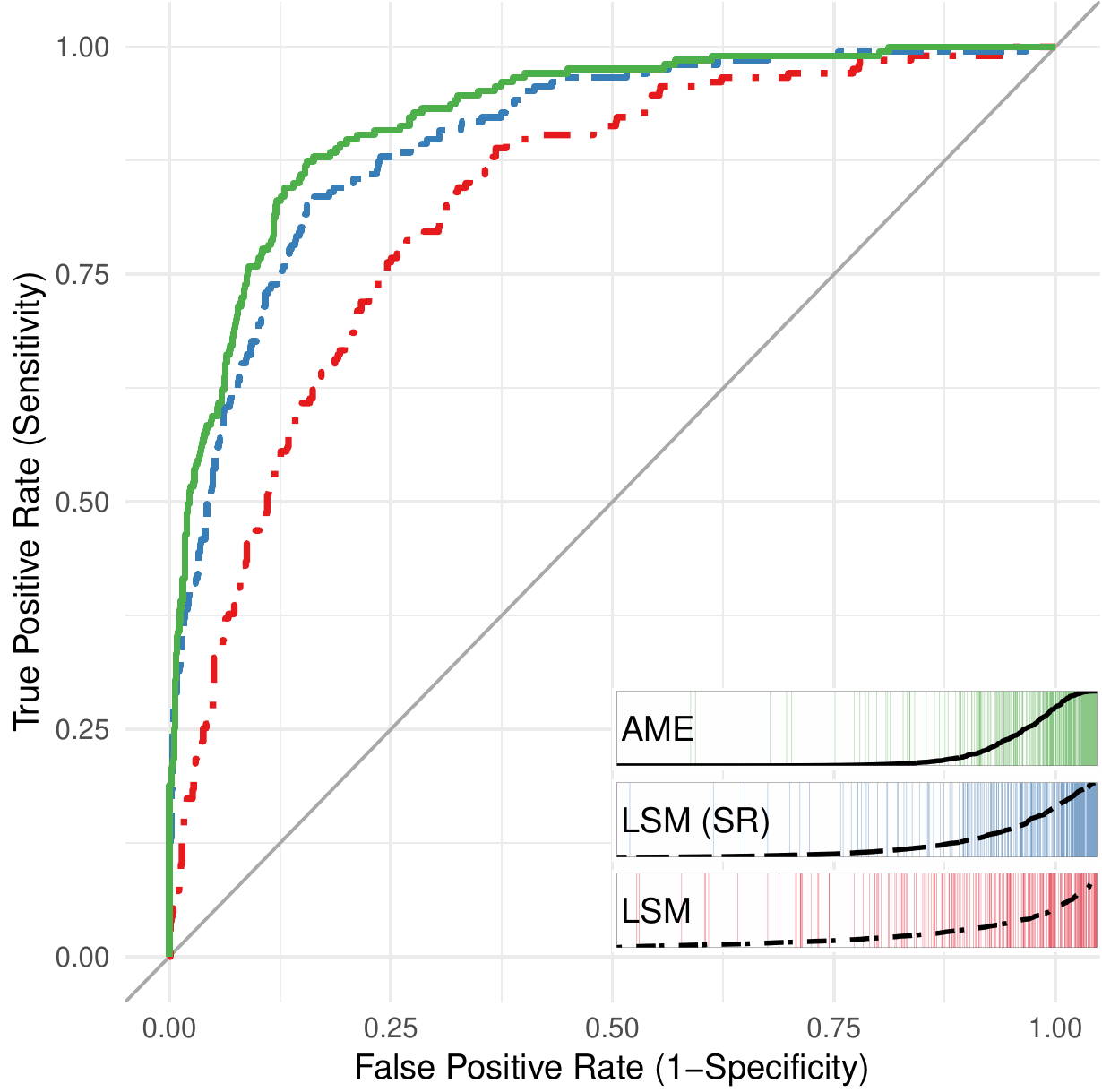} &
	\includegraphics[width=.5\textwidth]{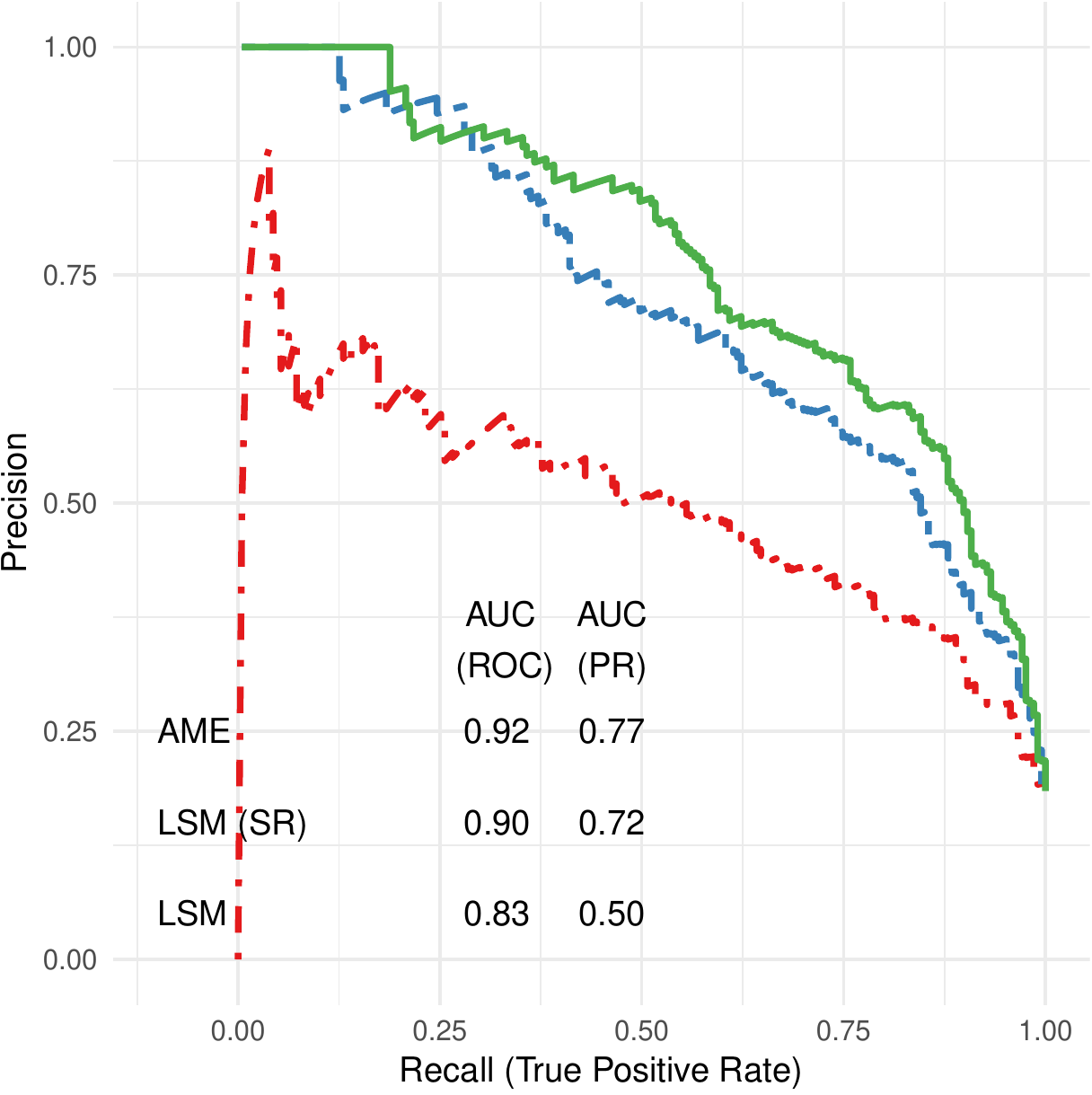}
	\end{tabular}
	\caption{Assessments of out-of-sample predictive performance using ROC curves, separation plots, and PR curves. AUC statistics are provided as well for both curves.}
	\label{fig:roc_latentSpace}
\end{figure}

\begin{figure}[ht]
	\centering
	\includegraphics[width=1\textwidth]{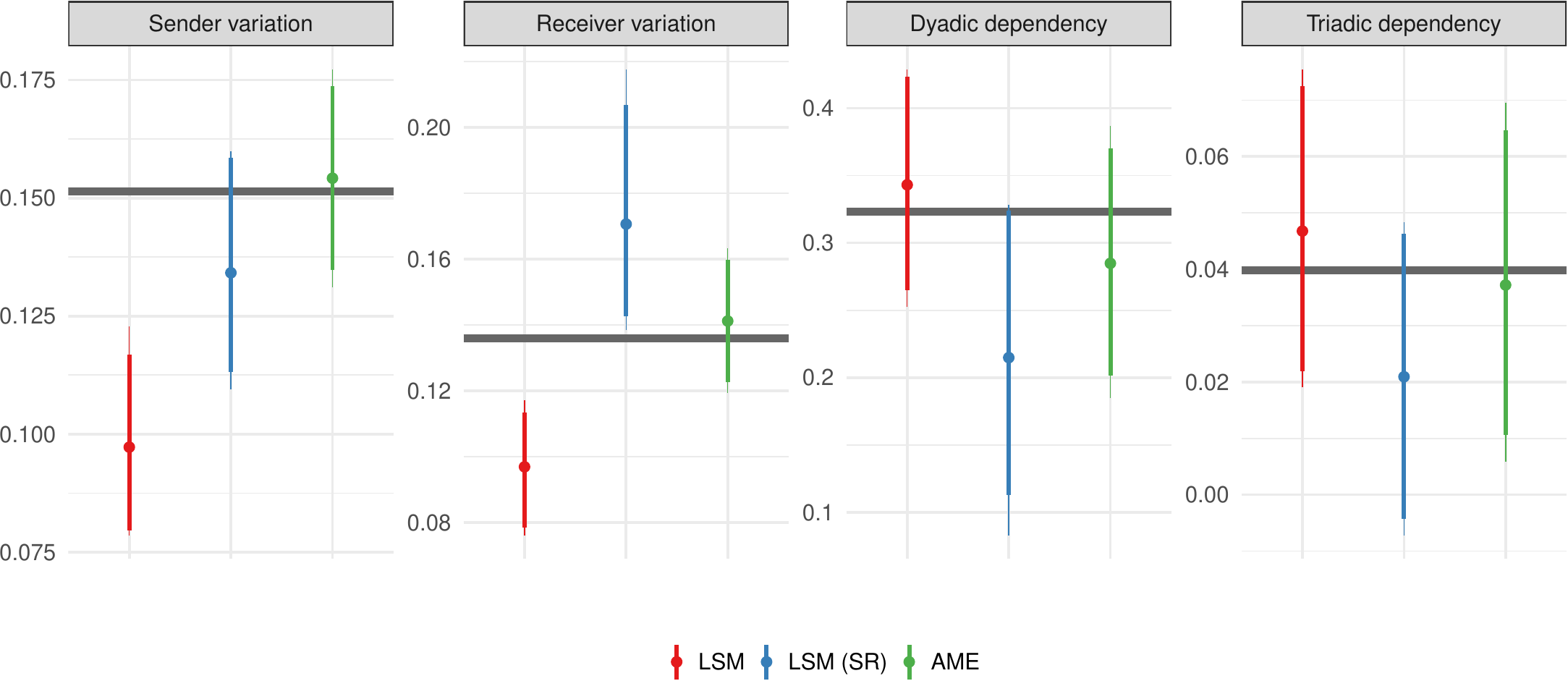}
	\caption{Network goodness of fit summary using \pkg{amen}.}
	\label{fig:netPerfCoef_latSpace}
\end{figure}

\FloatBarrier

\subsection*{Simulation Based Comparison of \pkg{amen} \& \pkg{latentnet}}

We construct a simulation study to examine differences in the ability of LSM and LFM to capture network dependencies under varying scenarios of ``egalitarianism''. By egalitarianism here we refer to how equally balanced the nodes are in terms of their number of ties. We construct six simulation scenarios representing varying degrees of egalitarianism. Note that to provide as fair a test as possible to the LSM we focus on comparing to just the LFM, the multiplicative effects portion of AME (see Equation 3 in the manuscript). This means that we exclude the additive effects described by the SRM portion of the model (see Equation 2 in the manuscript). 

For each scenario, we simulate fifty binary, directed networks with 100 nodes each and then evaluate the performance of LFM and LSM to predict this network structure. The results are shown in Figure~\ref{fig:sim_egal} below. Each panel here represents one scenario in which we vary the degree of egalitariansim. The left most panel represents the situation in which the structure of the network is most egalitarian. The numbers at the top of each panel indicate the standard deviation of the degree distribution averaged across fifty simulations. Across the diagnoal of the visualization, we also provide an example of the type of network that was simulated. The size of nodes in each example network corresponds to the number of ties that node has. As we go from left to right, we can see much greater variance in the size of nodes within the network, which indicates that the level of egalitarianism is changing. 

We run a LFM and LSM on each of the simulated networks from each scenario, and compare the predictive performance based on AUC (ROC) and AUC (PR) statistics. We set $K=2$ for both the LFM and LSM and estimate each model without any covariates. The results of this analysis indicate that under these varying scenarios of egalitarianism LFM consistently outperforms the LSM. However, the performance of both models tends to decline as the structure of the simulated networks become less egalitarian (i.e., the extent of tie formation among just a few nodes becomes much higher than the typical node in the network). If covariate information was provided to the model about which nodes were more likely to form ties, then the predictive performance of both models would obviously improve. Additionally, if we were to estimate the full AME model (SRM + LFM) then the additive effects would be able to capture the degree heterogeneity. Typically, in most applied scenarios one would always to include both the additive and multiplicative effects portions when using AME.\footnote{When using the \pkg{amen}, the SRM portion of the model will be included by default.} 

\begin{figure}[ht]
	\centering
	\includegraphics[width=1\textwidth]{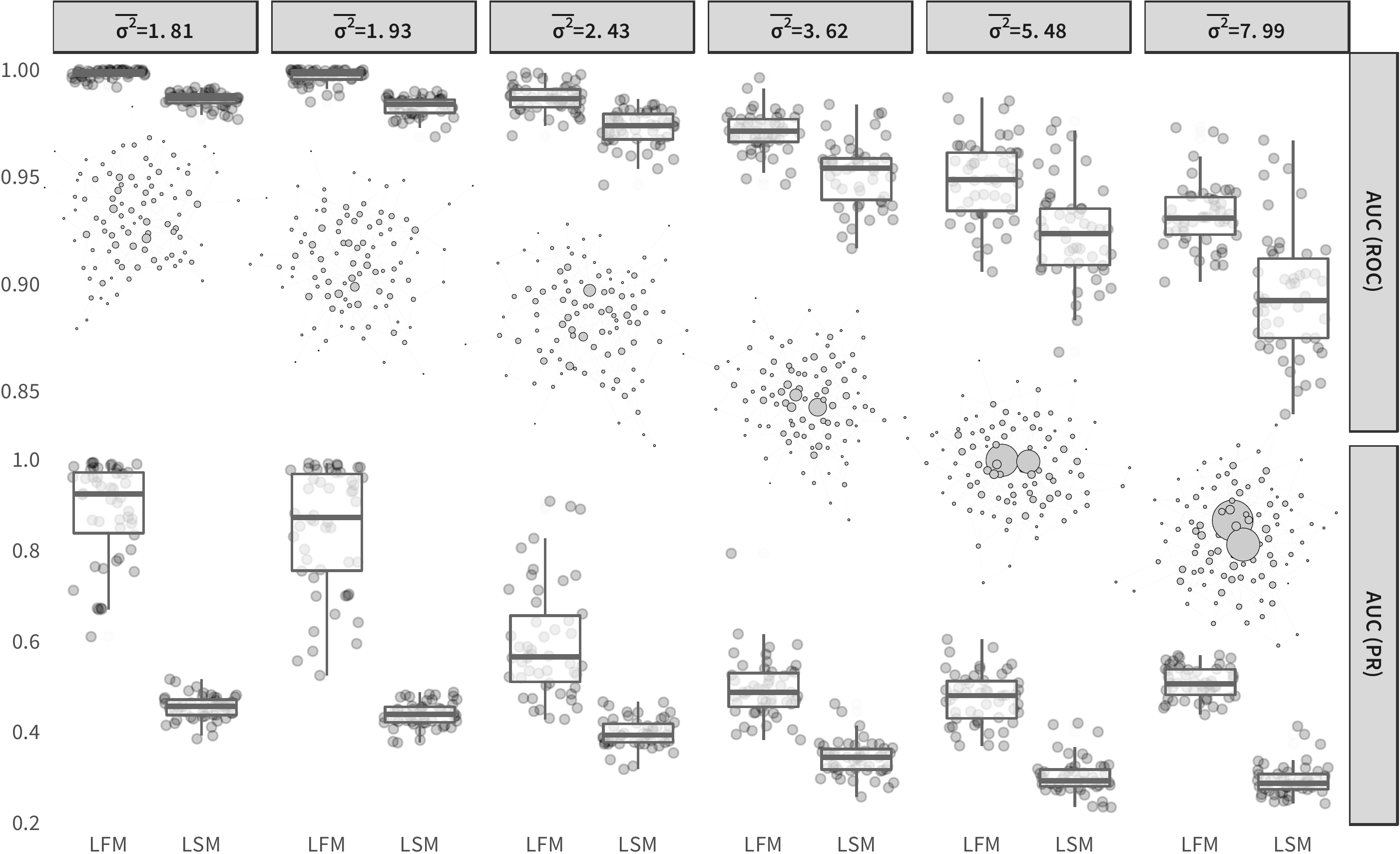}
	\caption{Predictive performance of LFM vs LSM for networks under five scenarios (the panels) that vary the extent to which the distribution of ties are egalitarian. We use a box plot to represent the performance of LFM and LSM across fifty simulations for each scenario. The set of network visualizations across the diagonal of the plot illustrate a representative network from one simulation under that scenario, and the size of nodes corresponds to their number of ties. The labels at the top of each panel indicate the standard deviation of the number of ties, which are averaged across the fifty simulations for that scenario.}
	\label{fig:sim_egal}	
\end{figure}
\FloatBarrier

\begin{figure}[ht]
	\centering
	\includegraphics[width=1\textwidth]{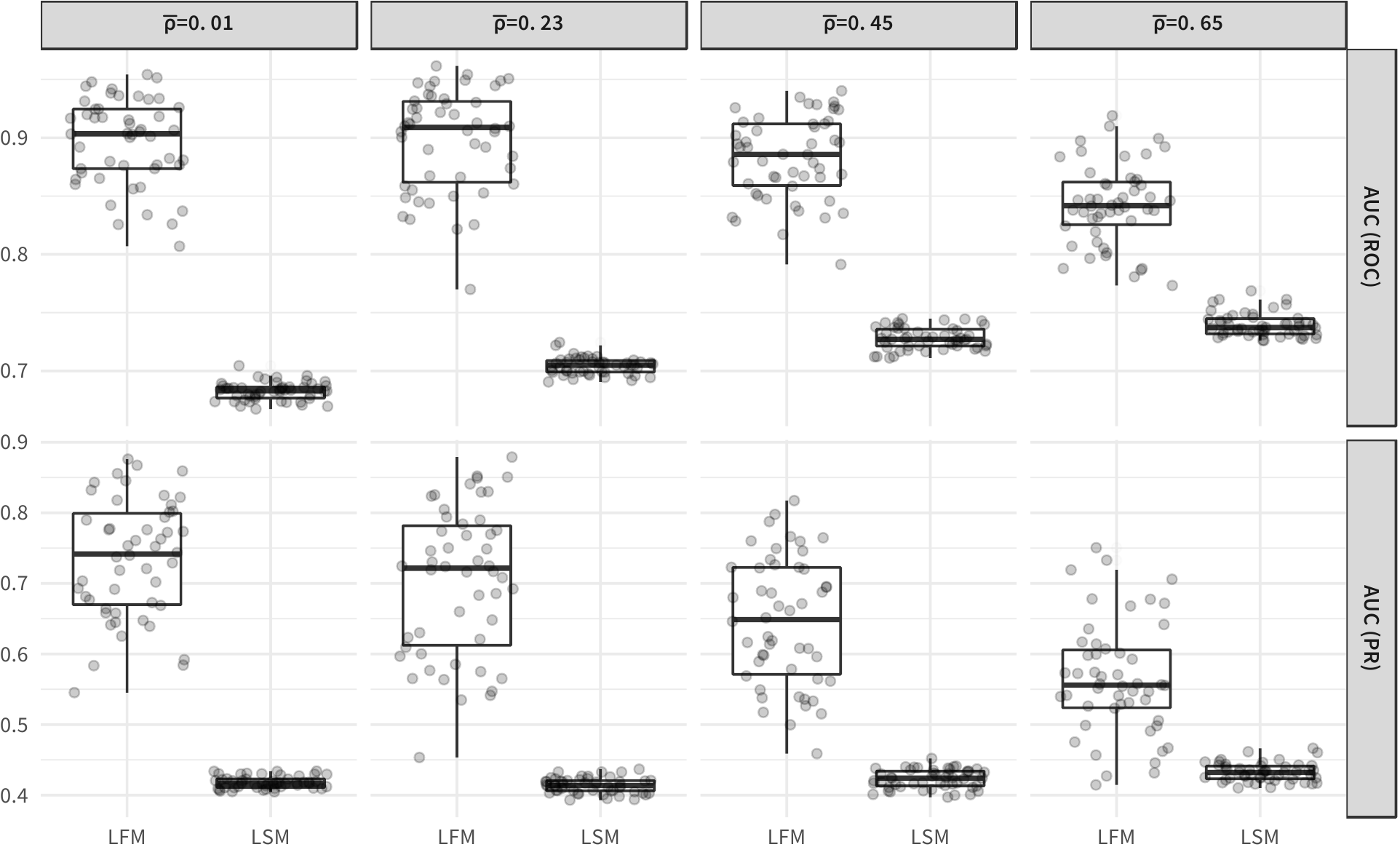}
	\caption{Predictive performance of LFM vs LSM for networks with varying levels of reciprocity. We use a box plot to represent the performance of LFM and LSM across fifty simulations for each scenario. The labels at the top of each panel indicate the average level of reciprocity across the fifty simulated in that scenario.}
	\label{fig:sim_recip}		
\end{figure}

Next, we construct a second simulation study to compare the predictive performance of LSM and LFM under varying levels of reciprocity. Here again we simulate a set of scenarios, and for each scenario we simulate fifty binary, directed networks with 100 nodes. The results are shown in Figure~\ref{fig:sim_recip}. Each panel here represents represents one scenario with a certain degree of reciprocity. The left-most panel highlights the case where there is little to no reciprocity in the network and the right-most where the level of reciprocity is quite high. The average level of reciprocity across the fifty simulated networks is given at the top of each panel. 

\FloatBarrier

To compare LFM and LSM, we again utilize AUC (ROC) and AUC (PR) statistics. $K$ is set to 2 for both models and no covariate information is provided. Here again we find that the LFM consistently outpeforms the LSM, though at higher levels of reciprocity the performance difference between the two approaches does shorten. If we were to estimate the full AME model, then we would be better able to capture reciprocity in the network, as dyadic reciprocity is estimated within the additive effects portion of AME. 

\clearpage
\bibliography{/Users/s7m/whistle/master}
\bibliographystyle{APSR}\biboptions{authoryear}
\newpage

\end{document}